\documentclass{article}

\usepackage{fullpage}
\usepackage{graphics}
\usepackage{graphicx}
\usepackage{amsmath}
\usepackage{natbib}
\usepackage{indentfirst}
\usepackage[utf8]{inputenc}
\usepackage[T1]{fontenc}
\usepackage{enumerate}
\usepackage{lineno} 
\usepackage{hyperref}
\usepackage{setspace}

\newcommand{\bm}[1]{\mbox{\boldmath $ #1 $}}

\begin{document}

\title{On Bayesian Modelling of the Uncertainties in Palaeoclimate Reconstruction}
\author{Andrew C. Parnell$^{1,2}$, James Sweeney$^{3}$, Thinh K. Doan$^{4}$, Michael Salter-Townshend$^{2}$,\\ Judy R.M. Allen$^{5}$, Brian Huntley$^{5}$, and John Haslett$^{4}$\\
\small $^{1}$School of Mathematical Sciences (Statistics), University College Dublin, Ireland \\
\small $^{2}$Complex and Adaptive Systems Laboratory, University College Dublin, Ireland \\
\small $^{3}$School of Mathematical Sciences, University College Cork, Ireland \\
\small $^{4}$School of Computer Science and Statistics, Trinity College Dublin, Ireland \\
\small $^{5}$School of Biological and Biomedical Sciences, Durham University, UK \\
}

\maketitle

\begin{abstract}
We outline a model and algorithm to perform inference on the palaeoclimate and palaeoclimate volatility from pollen proxy data. We use a novel multivariate non-linear non-Gaussian state space model consisting of an observation equation linking climate to proxy data and an evolution equation driving climate change over time. The link from climate to proxy data is defined by a pre-calibrated forward model, as developed in \cite{Salter-Townshend2012} and \cite{Sweeney2012}. Climatic change is represented by a temporally-uncertain Normal-Inverse Gaussian Levy process, being able to capture large jumps in multivariate climate whilst remaining temporally consistent. The pre-calibrated nature of the forward model allows us to cut feedback between the observation and evolution equations and thus integrate out the state variable entirely whilst making minimal simplifying assumptions. A key part of this approach is the creation of mixtures of marginal data posteriors representing the information obtained about climate 
from each individual time point. Our approach allows for an extremely efficient MCMC algorithm, which we demonstrate with a pollen core from Sluggan Bog, County Antrim, Northern Ireland.\\
\end{abstract}


\section{Introduction}\label{intro}

In this paper we aim to answer the following two questions:
\begin{quote}
(1) What was climate during period $(t_i,t_j)$ in location $s$?\\
(2) What was the \textit{volatility} in climate during period $(t_i,t_j)$ in location $s$?
\end{quote}
For illustration, we set $(t_i,t_j)=(0,14)$ thousand years before present (written `ka BP'\footnote{We use the common convention in setting the `present' to be 1950AD}), and $s$ to be the central part of Northern Ireland. Were $t_j$ to be restricted to a smaller value we could use recent instrumental measurements for which we could obtain a fairly precise answer. When $t_j$ is allowed to grow large, we need to use proxy data: indirect records of climate stored in biological and mineral deposits. In this paper we outline a method to perform inference on the palaeoclimate (sometimes loosely referred to as `reconstruction') and thus answer our questions of interest using pollen data, following on from \cite{haslettetal2006}. However, our methods are general and readily apply to other proxy data series or even multiple proxies simultaneously.\\

Palaeoclimate reconstruction is a major focus of the Intergovernmental Panel on Climate Change \citep{ipcc2007ch6}. Public interest, however, has largely been fuelled by the `Hockey Stick' and `climategate' controversies, e.g. \citet{Mann1998,Mann1999,McShane2011} where $(t_i,t_j)$ was $(0,1)$ka BP and $s$ was the Northern Hemisphere. Climate changes during the past millennium are relatively small and can be inferred with reasonable precision from proxies (e.g. tree rings) that are resolved annually. In contrast, the much older Younger Dryas period (12.8ka to 11.5ka BP) shows a rapid switching from warm to cold to warm. During this period ice core data from Greenland show abrupt warmings of up to $16\,^{\circ}\mathrm{C}$ within decades \citep[][P435]{ipcc2007ch6}. This size and rate of change is not captured well by the General Circulation Models (GCMs) which are used to predict future climate, nor by the precise proxies used to examine the past millennium. Pollen proxy data offer the best hope of 
resolving such sizeable past climate changes in locations other than Greenland.\\

The model we propose differs from many recent studies performing inference on palaeoclimate because (a) it is non-linear and non-Gaussian in the relationship between climate and proxy, (b) we infer only climatic variables to which the proxy is sensitive, (c) we use real data to produce climate reconstructions rather than simulated `psuedo-proxies', (d) we take account of chronological uncertainty, and (e) we allow for stochastic variability in climate. This approach, tackling many more sources of uncertainty than previously, is inspired by the SUPRANET group\footnote{Studying uncertainty in palaeoclimate reconstructions: http://caitlin-buck.staff.shef.ac.uk/SUPRAnet/}, of which many ideas are shared with \cite{Tingley2012} and \cite{Huntley2012}. We expand more on the importance of these various aspects in subsequent sections.\\

The pollen data we use arise from cores taken beneath a lake or bog. The chosen site will have accumulated many thousands of years of pollen from plants and trees that have surrounded it. After extraction, the core will have been sliced into narrow depths $d_i$, each treated as a near instantaneous snapshot of the vegetation at that depth. From each slice a palynologist will count many different varieties of pollen and record the counts and depths. We make use of 28 of these counts, corresponding to pollen varieties that have been shown to be representative of a particular aspect or boundary condition on one or more aspects of climate \citep{huntley1993}. We write this 28-vector as $y_i$. The information about climate from the pollen is obscured by: the delay in the reaction of plants and trees to climate change; the ability of the palynologist to recognise the desecrated remains of fossil pollen under a microscope; and the degree to which different species disseminate differentially recognisable pollen. The 
former of these is considered negligible over the timescales with which we work. The latter two are included as part of our likelihood, being both zero-inflated and highly non-linear in the relationship between climate and pollen.\\

The practicalities of sampling lake sediment mean that the depths $d_i$ may not be taken regularly throughout the core. Besides, even if the depths in the core are regularly taken, the times to which they correspond depend on the rate of sedimentation and may be very non-linear. The usual approach to transform depth into time is to take between 5 and 20 radiocarbon dates from various depths (again not necessarily regularly), and use these to build a monotonic age-depth model which supplies times $t_i$ from every layer in the core; a number of statistical age-depth models have been proposed \citep[e.g.][]{Ramsey2008,haslettparnell2008b,Blaauw2011}. A side-effect of using one of these models is that times $t_i$ are now uncertain, governed by a draw from a probability distribution $\pi(t|d)$ with $t$ and $d$ representing all of the times and depths in the core. Each draw is termed a chronology. \\

Taken altogether, we can write our model in state space form as:
\begin{eqnarray}
\label{statespace1}
y_i|c_i \sim f_\theta(c_i,\mathcal{D}),\; i=1\ldots,n\\ 
\label{statespace2}
c_i = c_{i-1} + \gamma_i, \; i=2,\ldots,n
\end{eqnarray}
where $y_i=y(t_i)$ represents pollen data at time $t_i$, $c_i=c(t_i)$ is palaeoclimate, $f$ is a \textit{forward model} parameterised by $\theta$, and $\gamma_i \sim N(0,v(t_{i-1},t_i))$ are residuals. Following convention, we call Equation \ref{statespace1} the \textit{observation equation}, and Equation \ref{statespace2} the \textit{evolution equation}. All of the parameters involved are multivariate: $y_i$ being here of dimension 28 and $c_i$ being of dimension 3. In our example in Section \ref{results} $n=115$. The parameters $\theta$ and the relationship $f$ are well known and thus pre-calibrated from external data sources (a modern data set of pollen and climate $\mathcal{D} = (c^m,y^m)$), and so our focus is on the posterior distribution $\pi(c,v|y)$. Our particular interest is in the distribution of the incremental variance terms $v_i=v(t_{i-1},t_i)$, representing the square of the volatility. We set these as being Inverse Gaussian \citep{Barndorff-Nielsen1997}, written $IG(\eta,\phi)$, yielding an infinitely divisible long-
tailed stochastic process on $c$. \\

From the age-depth model and the state space equations above we can form a posterior distribution:
\begin{eqnarray}
\label{eqn1}
\pi(c,v,t,\eta,\phi|y,d) \propto \pi(\eta,\phi)\; \pi(t|d) \;\pi(c_1) \; \prod_{i=1}^{n-1} \pi(v_i|\eta,\phi,t) \; \prod_{i=2}^n \pi(c_i|c_{i-1},v_i) \;  \prod_{i=1}^n \pi(y_i|c_i)
\end{eqnarray}
where $\pi(\eta,\phi)$ is the prior distribution on $(\eta,\phi)$, $\pi(t|d)$ is the age-depth model, $\pi(c_1)$ is the marginal prior on the first climate time point (which we treat as flat), $\pi(v_i|\eta,\phi,t)$ is the Inverse Gaussian prior on increment variances, $\pi(c_i|c_{i-1},v_i)$ is the evolution equation distribution, and $\pi(y_i|c_i)$ is the likelihood from the observation equation distribution.\\

\begin{figure}[!h]
\begin{center}
\includegraphics[width=14cm]{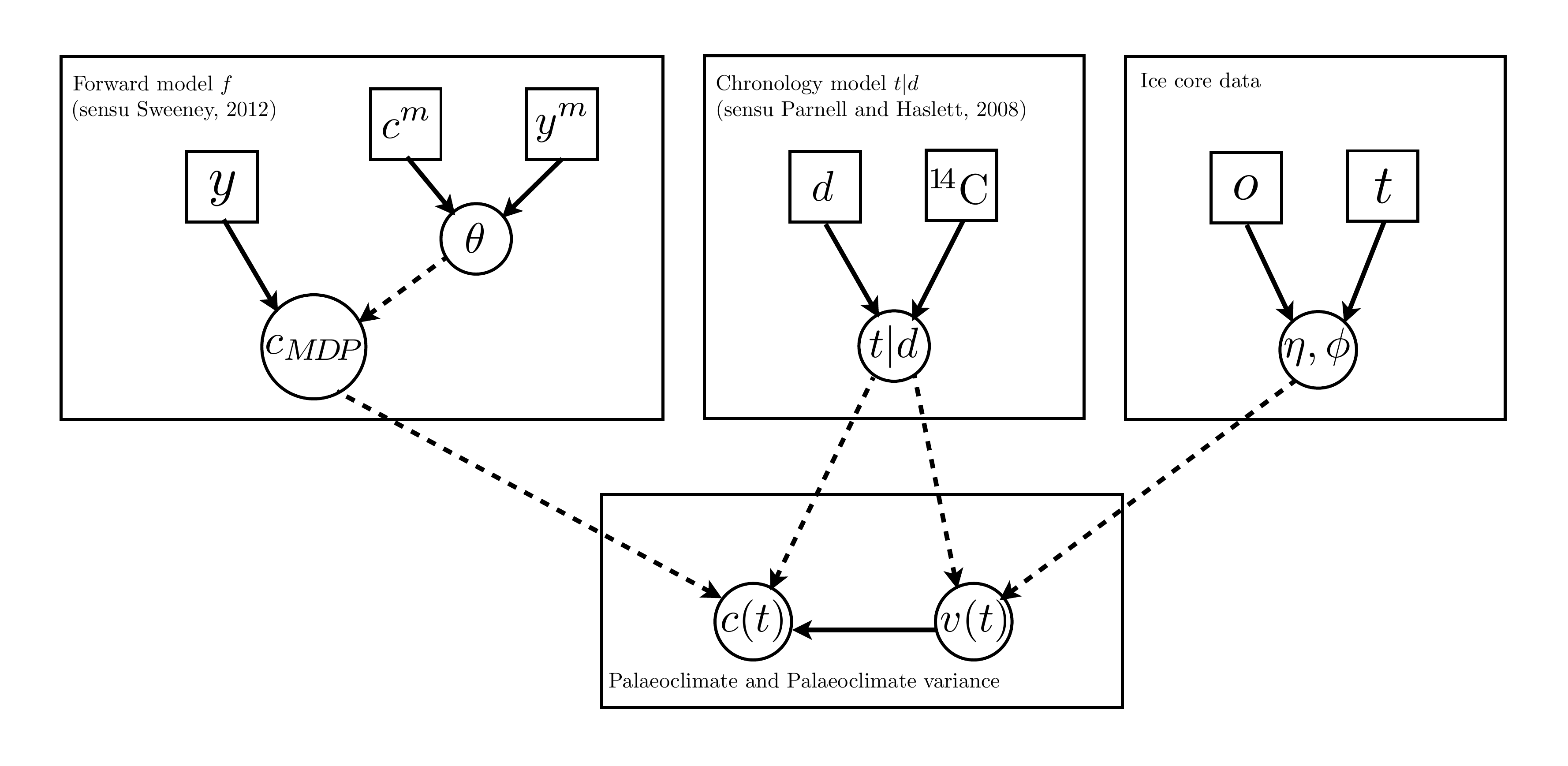}
\end{center}
\caption{
A Directed Acyclic Graph for our palaeoclimate model. Broken lines indicate relationships where feedback is cut. See Section \ref{model} for more details on the modelling relationships and Technical Appendix \ref{notation} for details on the notation.}
\label{DAG}
\end{figure}

Our focus in this paper is not on the likelihood $\pi(y_i|c_i)$, the version we use having been discussed in detail by \cite{Salter-Townshend2012}, and \cite{Sweeney2012}. Instead, we focus on how to fit the model defined by Equation \ref{eqn1} in a situation where repeated calls to the likelihood are costly. In fact we separate out the calculation of the likelihood into:
\begin{enumerate}[(a)]
\item An initial calibration step, where we cut feedback from $\theta$ to form $\pi(y_i|c_i) = \int \pi(y_i|c_i,\theta) \pi(\theta|\mathcal{D})\;d\theta$. From these we create \textit{marginal data posteriors} (MDPs) of the form $\pi(c_i|y_i)$. 
\item A second model-fitting step where we use the MDP in place of the likelihood to create the posterior distribution  $\pi(c,v,t,\eta,\phi|y,d)$ as above. 
\end{enumerate}
We justify the validity of step (a) in Section \ref{MDPs}; the remainder of the paper is concerned with step (b). A Directed Acyclic Graph (DAG) for our complete model is shown in Figure \ref{DAG} and discussed further in Section \ref{model}.\\

A key aspect of our approach is that many of the parts of Equation \ref{eqn1} are treated as mixtures of Gaussian distributions. The marginal data posteriors are approximated as finite mixtures of Gaussians, which allows the posterior to be factorised, and removes the need for $c$ to be estimated during the model fitting stage. We can then create sample paths of $c(t)$ as a mixture of Gaussian distributions depending on the NIG parameters and the MDPs. Similarly, the NIG prior can be seen as a mean-scale infinite mixture of a Gaussian distribution where the mixing distribution is Inverse Gaussian. We make use of this property to give fast updates of the increment variance parameters $v$.\\

Our approach fundamentally differs from previous work in the use and formatting of data used as `input' to an inference procedure. The work of \citet{Mann1998} and \citet{Mann1999} use `products': pre-treated proxy data sets that have been aggregated and transformed on to a scale that is supposed to match an aspect of climate such as mean annual temperature. Often the details of such transformations are lost to the final inference so that the link between proxy data and climate, especially with respect to uncertainty, is not considered during the reconstruction of past climate. Similarly, other authors \citep[e.g.][]{Li2010a} have simulated their proxy data from climate model output, so that the input format is analogous to that of products. By contrast, we use an inference procedure that takes raw pollen data directly to perform inference on climate. However, the marginal data posteriors we produce might be considered as products though with a more fully quantified estimate of uncertainty.\\

The paper is organised as follows. In Section \ref{previous} we examine previously used methods for creating palaeoclimate reconstructions. In Section \ref{motivating} we present a simple version of our time series model based on ice core data. In Section \ref{MDPs} we introduce the mathematical justification of our method, and apply it to the palaeoclimate scenario in Section \ref{model}. In Section \ref{results} we show the results of our model applied to a site in Northern Ireland. We examine some shortcomings of our model and discuss future directions in Section \ref{discuss}. The mathematical derivation is somewhat complex so we include a notational glossary in Technical Appendix \ref{notation}. We include some of the more technical details of the algorithm in Technical Appendix \ref{MCMC}. We perform model validation checks in Appendix \ref{validation}. The resulting model is implemented in the R package Bclim, which is available online at the R repository \citep{
RFoundationForStatisticalComputingAustria2011}. Instructions for using the R package can be found in Technical Appendix \ref{Bclim}.

\section{Previous work in statistical palaeoclimatology}\label{previous}

Here we explore the ways in which other authors have dealt with various aspects of palaeoclimate reconstruction. Whilst recent work has focussed on model-based approaches (particularly Bayesian), authors in the more distant past have used other methods. The most famous of these is the \citet{Mann1998,Mann1999} papers using principal components regression or the regularised EM algorithm, though others such as \cite{huntley1993} and \cite{TerBraak1995} follow differing non-model based criteria. Other projects have focussed on certain aspects of palaeoclimate in relation to General Circulation Models (GCMs) or at particular points in time, such as PMIP \citep{Braconnot2003}, and the PALAEOQUMP project \citep{Masson-Delmotte2006}. For brevity, we discuss in detail here only the statistical model-based approaches used by \citet{haslettetal2006, Tingley2010, Li2010a} (hereafter H06, T10 and L10 respectively) and focus on the key climatological issues: choice of climate variables, likelihood choices, prior 
distributions on climate, chronological modelling; and the most relevant statistical issues: fitting models with complex likelihoods, the NIG process, and the treatment of mixture models. A more thorough review of different approaches can be found in \cite{Tingley2012}. \\

\subsection{Issues of climatology} 

A key modelling choice in palaeoclimate reconstruction is the aspects of climate to reconstruct; climate being defined as the `average' of weather. Many authors \citep[e.g.][L10]{Mann1998} have reconstructed mean annual temperatures, often over large spatial scales such as the Northern Hemisphere. This may seem like a logical choice, given that these are relatively easy to measure and relevant to human wellbeing. However, as recently summarised by \citet{Huntley2012}, the choice should instead depend on the sensitivity of the proxy data to changes in climate. Many biological proxies have shown that they are not sensitive to changes in mean annual temperature so reconstructions of this sort will be biased and overly uncertain. Instead, more relevant variables were proposed by, amongst others, \citet{huntley1993}:
\begin{itemize}
\item the Mean Temperature of the Coldest Month (MTCO), a measure of the harshness of the winter,
\item the Growing Degree Days above $5^{\circ}\mathrm{C}$ (GDD5, also known as the annual temperature sum above $5^{\circ}\mathrm{C}$), a measure of the length of the growing season,
\item the ratio of Actual to Potential Evapotranspiration (AET/PET), a measure of the proportion of available moisture.
\end{itemize}
The first two of these were used for reconstruction by H06. Even then, however, missing a single important climate variable (in this case concerning moisture) will increase uncertainty, bias and possibly invalidate the conditional independence of the likelihood. Other non-biological proxies \citep[e.g. Borehole temperatures,][]{Brynjarsdottir2011} have shown sensitivity to surface temperature of only the surrounding microclimate. The choice of climate variables is even more important in multi-proxy reconstructions, where the differing sensitivities and temporal scales at which proxies are formed needs to be carefully understood before models can be created.\\

Once the climate variables have been chosen, we can posit a likelihood $\pi(y|c,\theta)$. We see the likelihood here as a data generating mechanism providing realistic proxy data given the relevant aspects of climate, often also called a forward model indicating the direction of causation (from climate \textit{to} proxy). Numerous methods for creating a link between climate and pollen proxy data have been suggested, see \cite{ohlwein2012} for a full review. There are two approaches which stand out as being most popular. The first is to perform a calibration via modern analogue data, as used by H06, T10 and L10. The idea is to find data, here $\mathcal{D}$, for which both precise proxy and climate are available, and upon which the likelihood can be trained. The version used by T10, L10 requires the calibration data to be temporally overlapping, through which a temporal proxy/climate  relationship might be inferred. The version used by H06 has no temporal overlap, but rather uses the climatological location of 
modern data in a multivariate psuedo-spatial model where the proxy is the response. This model has been substantially enhanced in \cite{Salter-Townshend2012} and \cite{Sweeney2012}, and the latter is the version we use for our palaeoclimate model. The second approach is to use a physical model, as employed by \cite{Garreta2009} with mixed success. The physical model can provide a far richer relationship, taking account of differences in pollination mechanism and productivity between trees, and their response to complex inter-relations between climate variables. However, the models used are often extremely slow and may lack a stochastic element, making their use in Equation \ref{eqn1} difficult. \\

In either scenario a likelihood is created such that:
\begin{eqnarray}
y_i|c_i,\theta \sim f(h(c_i,\theta_h),\theta_\epsilon)
\end{eqnarray}
where $h$ is a function which accounts for the sensitivity of the proxy to climate with parameters $\theta_h$, and measurement error captured by $\theta_\epsilon$. Together $\theta=(\theta_h,\theta_\epsilon)$. Both parameters and their relationship through $h$ are estimated from the calibration data, $\mathcal{D}$. In \cite{Sweeney2012} this relationship is explicit; $f$ is a zero-inflated Negative Binomial distribution, with $\theta_h$ the parameters of a  Gaussian Markov Random Field within a nested Dirichlet distribution specifying the proportions of each taxa. In L10 and T10 $f$ is Gaussian (the proxy data $y$ are `products') with $h$ a linear regression so that $\theta_h$ represents regression parameters and $\theta_\epsilon$ residual variances and covariances. Both L10 and T10 use simulated data (`pseudo proxies') so it is not stated exactly which transformation of $y$ might link back to a real-world biological proxy such as pollen. \\


In H06 the prior distribution on climate was a $t_8$ random walk which is not closed under addition. Both L10 and T10 use more conventional time series methods, employing a Gaussian MA(2) and AR(2) model respectively. In T10 the prior distribution is allowed to vary spatially with a stationary Exponential covariance structure. In L10 the prior is expanded to include extra covariates corresponding to known forcings in a linear fashion. Unfortunately these structures are not applicable to the present situation where time is uncertain and continuous. In palaeoclimate reconstruction from pollen, uncertainty in time arises from radiocarbon dates taken from the pollen core which require non-linear calibration \citep[see][]{bucketal1996}. These contrast with the other proxy data used by L10 and T10 such as tree rings for which the times are known precisely. For our purposes we use the Bchron software of \citet{haslettparnell2008b}, being explicitly built for inclusion in palaeoclimate reconstructions.

\subsection{Issues of statistical modelling}

When the likelihood and prior distributions are Gaussian, as in T10 and L10, then Equation \ref{eqn1} is conceptually easy to solve via traditional MCMC or Particle-type methods. Things are complicated slightly by the high-dimensionality of the climate and volatility parameters and the sheer quantity of data. H06, however, have a non-linear model and thus combine the MCMC methods with some approximations which cut feedback between parts of the model. In particular, parameters $\theta$ of the likelihood are learnt solely from the modern data. Thus information from any particular core contributes no `further' information. In this paper we use a likelihood which is considerably more complex than that of H06, and is thus far too slow to solve using these methods.\\

A number of alternatives have been proposed for fitting models with complex likelihoods, including but not limited to: Approximate Bayesian Computation \citep[ABC;][]{Beaumont2002}, which relies solely on simulations from the likelihood and user-specified sufficient statistics to compute likelihood `distances'; and statistical emulation \citep[SE;][]{kennedy2001}, which builds a stochastic approximation to the likelihood through a designed experiment. However, since we have reasonably good access to a well-defined likelihood it seems inappropriate to throw away much of the information in an ABC type algorithm. Similarly, whilst SE has been shown to work well in a high dimension climatological situation \citep{Holden2010}, it is mostly used for deterministic models \citep[though see][]{Henderson2010}.\\

%

As stated earlier, our model can be usefully seen in multivariate state-space terms for which there are numerous algorithms based on Sequential Monte Carlo \cite[see, e.g.][for a review]{Carvalho2010}. These proceed (in our notation) in a `forward' or filtering stage\footnote{Unrelated to the forward model $f$ presented in Equation \ref{statespace1}.} by simulating from $\pi(c_1)$ and then forming $\pi(c_k|y_1,\ldots,y_k)$ sequentially for $k=2,\ldots,n$. The filtering densities $\pi(c_k|y_1,\ldots,y_k)$ are related to our marginal data posteriors, being posterior distributions on a subset of the data. However, their creation requires both the use of the evolution and the state equation for every time point. In the subsequent sections, we show that in our situation it is possible to produce a valid joint posterior via only the MDPs, which does not require calculation of the evolution density at the forward stage.\\

A related problem in the estimation of state space models is known as the smoothing or `backwards' step to produce a full joint posterior $\pi(c_1,\ldots,c_n|y_1,\ldots,y_n)$. It is well-known that this is a challenging problem for particle methods \citep{Doucet2008}, especially in the case of large $n$ and where inference on other parameters are also required. Indeed, our focus is on this smoothing stage of state-space modelling, though even this step (when using the nomenclature of particle filters) over-stresses the focus on the posterior mean and is less natural when the focus is, as here, on past volatility. By contrast, we are in a somewhat unusual situation in that we are happy to cut feedback so that the likelihood depends only on the state parameters (climate $c$) with the distribution on $\theta$ known. When the prior is intrinsic we show that it is possible to integrate out our state variable $c$ and bypass many of the issues caused by particle degeneracy and sample size. We thus believe that our 
MDP approach is superior (achieving faster convergence even for large $n$), though we report elsewhere on the general properties of our fitting algorithm.\\

We use the NIG as a prior distribution on differences $c_i-c_{i-1}$ across continuous time. The distribution was introduced by \cite{Barndorff-Nielsen1997} as a mean-scale mixture of Gaussian distributions. If $x_i|v_i \sim N(\mu + \beta v_i, v_i)$ and $v_i \sim IG(\alpha,\delta)$ where $IG$ is in the Inverse Gaussian distribution \citep{Betro1991}, then $x_i \sim NIG(\mu,\beta,\alpha,\delta)$ with mean $\mu$, skew $\beta$, and variance/tail parameters $\alpha,\delta$. Most importantly, the NIG distribution is closed under addition, so that if $X_i \sim NIG(\mu_i,\beta,\alpha,\delta_i)$ then $\sum_i X_i \sim NIG(\sum \mu_i,\beta, \alpha,\sum \delta_i)$. Bayesian inference for the NIG distribution has been discussed by \cite{Karlis2004}, providing closed-form complete conditionals and thus a neat Gibbs algorithm. Some of these complete conditionals transfer over to our model and are discussed further in Section \ref{motivating}. The form we use is that of the NIG process as discussed by \cite{Ribeiro2003}, 
representing the random walk as a subordinated Brownian motion. The resulting NIG process is, of course, not the only long-tailed smoothing model available, forming part of the larger class of Levy processes \citep[e.g.][]{Barndorff-Nielsen2001}. Similarly, \cite{Fonseca2011} provide a class of long-tailed smoothers for space-time models. Where discontinuities are expected, a change-point model might be used \citep[e.g][]{Wyse2010}. However, the NIG process is extremely simple to work with, and provides many of the features we might expect to appear in a dynamical system such as climate.\\

The finite mixture models of which we make use are taken from the Mclust algorithm inspired by \citet{Fraley2002}. Here, a multivariate random variable $x$ arises as a mixture of $G$ normal distributions with unknown means and variances, so that $x = \sum_{g=1}^G p_g N(\mu_g,\tau_g)$ where the $p_g$ are mixture proportions. We use this technique to approximate our MDPs as mixtures of Gaussians via the EM algorithm \citep{Dempster1977} which provides posterior modes for $\mu_g,\tau_g$. A full summary of the methods by which this type of model may be fitted can be found in \cite{Fruhwirth-Schnatter2006}. An infinite mixture version via the Dirichlet Process \citep[e.g.][]{Congdon2001} would also be appropriate were we to be interested in simultaneously estimating the number of mixture components $G$.

\section{A motivating example: the GISP 2 ice core}\label{motivating}

\begin{figure}[!p]
\begin{center}
\includegraphics[width=14cm]{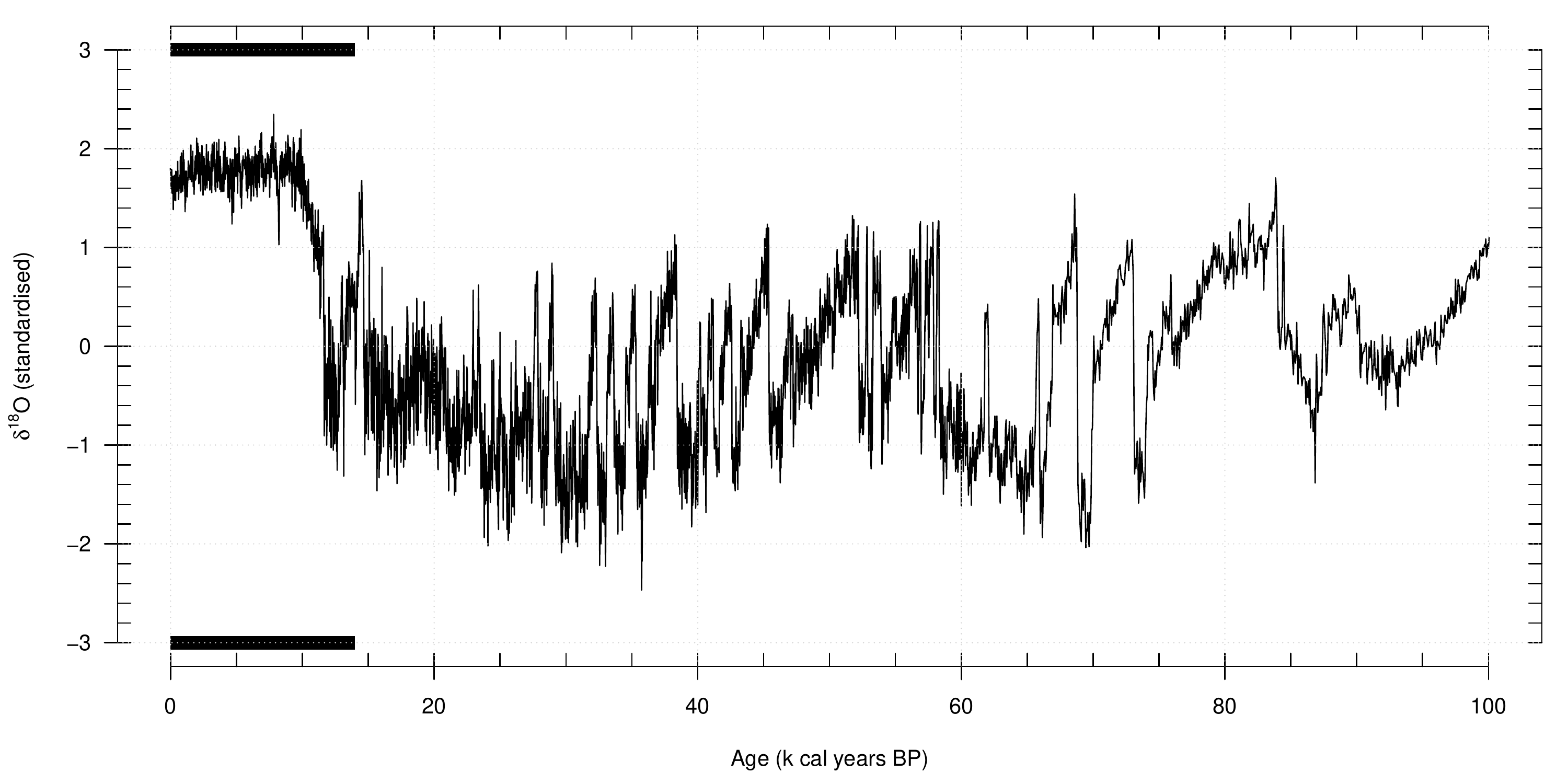}
\end{center}
\caption{
The GISP2 ice core data of \citet{Stuiver2000}. The $y$-axis contains the stable isotope measurement $\delta^{18}$O after standardisation to have mean 0 and standard deviation 1. A high value corresponds approximately to warmer summer temperatures, and the lower to colder summer temperatures. The period 0 to 14ka BP is highlighted to match with Figure \ref{icecorepost} (right panel) which shows the volatilities from this period.}
\label{icecore}
\end{figure}

\begin{figure}[!p]
\begin{center}
\begin{tabular}{cc}
\includegraphics[width=7cm]{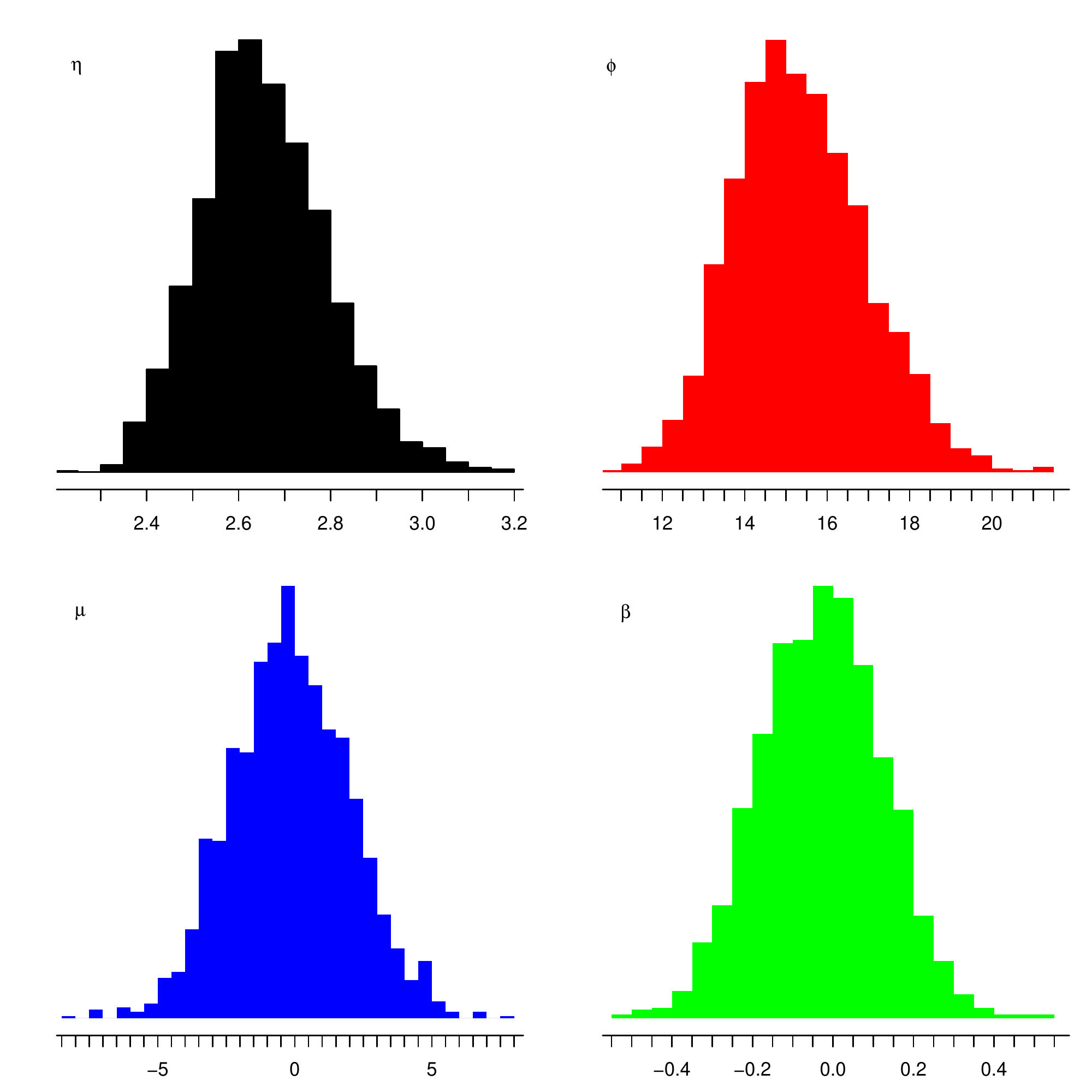} & \includegraphics[width=9cm]{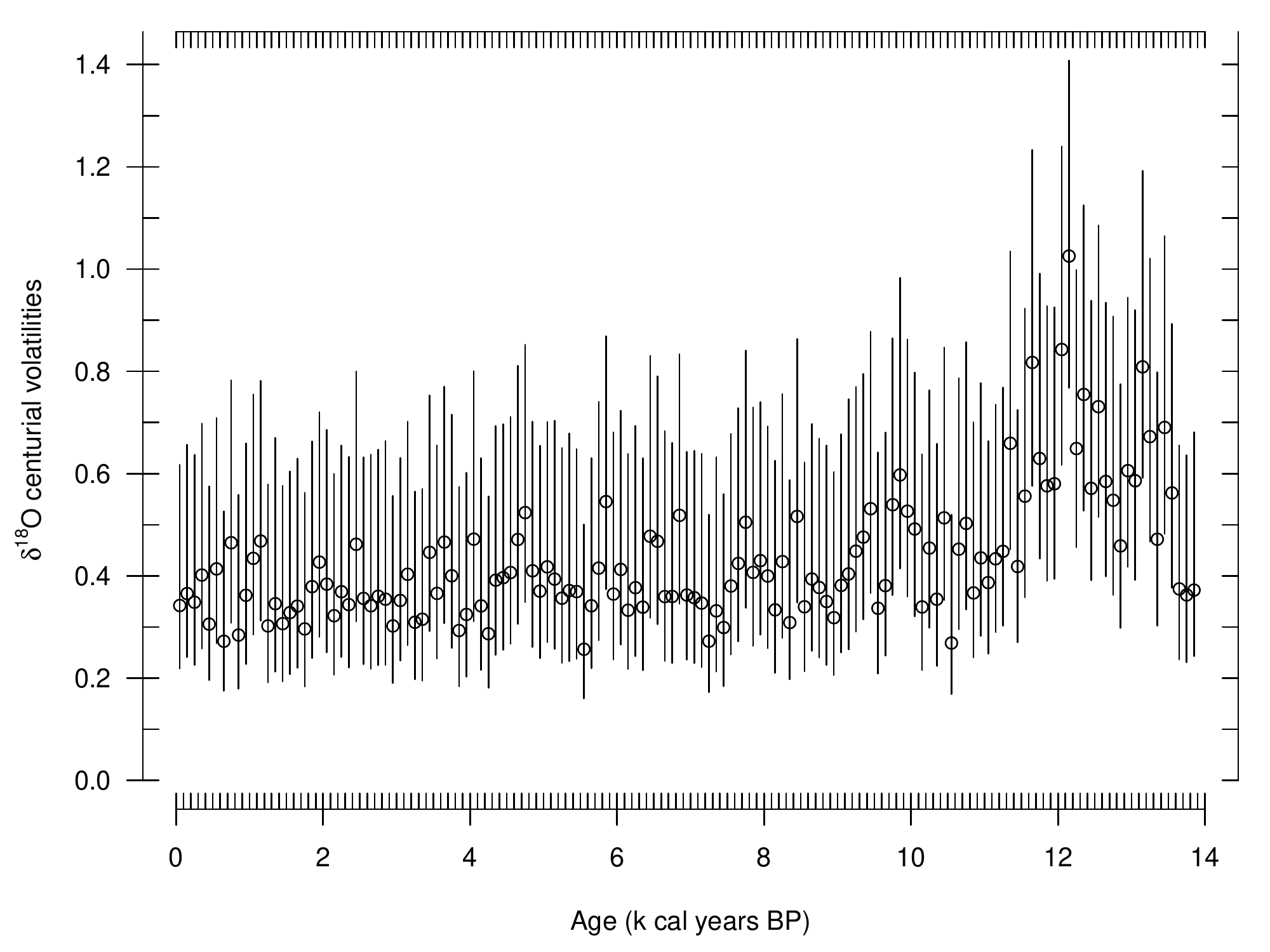}
\end{tabular}
\end{center}
\caption{
Posterior plots from fitting the model defined by Equation \ref{icecoremodel} to the GISP2 ice core data. The left plots show the posterior distributions of parameters $\mu$, $\beta$, $\eta$ and $\phi$, whilst the right-most plot shows the posterior distribution of $v$ over the time interval 0 to 14ka BP. The empty circles indicate the posterior mean whilst the vertical lines indicate the 95\% credibility intervals. The century with the highest (posterior mean) climate volatility is 12.1k to 12.2k BP.}
\label{icecorepost}
\end{figure}

We illustrate our long-tailed smoothing approach and how we might answer our motivating questions in a simplified scenario based on the GISP2 ice core of \citet{Stuiver2000}. A similar but more sophisticated analysis has been undertaken by \citet{Davidsen2010}. The proxy data here are measurements of $\delta^{18}$O, the deviation from a standard of the ratio of the two stable isotopes of oxygen measured in laminated (layered), and thus precisely dated, ice. However, the data are not observed at regular time intervals. For illustration we assume an identity relationship between $\delta^{18}$O and summer temperatures may be appropriate. A plot of the data covering the previous 120ka BP is shown in Figure \ref{icecore}. We write the response variable here as $o_i=o(t_i)$, and fit a model so that:
\begin{eqnarray}
\label{icecoremodel}
o_i-o_{i-1} \sim N(\mu \Delta_i + \beta v_i, v_i),\; v_i \sim IG_2(\eta \Delta_i,\phi \Delta_i),\; i=2,\ldots,n
\end{eqnarray}
where $\Delta_i=t_i-t_{i-1}$ and $IG_2$ is the alternative parameterisation of the Inverse Gaussian distribution given by \citet{Betro1991}. Under this parameterisation, the parameter $\eta$ can be seen as the  mean unit-time variance $v$ of the resulting NIG distribution, with parameter $\phi$ controlling the variance of $v$. Following \cite{Karlis2004}, we set prior distributions so that $\beta \sim N(0,\tau_\beta^{-1})$, $\mu \sim N(0,\tau_\mu^{-1})$, $\eta\sim Ga(a_\eta,b_\eta)$ and $\phi \sim Ga(a_\phi,b_\phi)$, where $Ga$ is a Gamma distribution. An MCMC algorithm now has Gibbs' steps with the following complete conditionals:
\begin{eqnarray*}
\pi(\eta|\ldots) &\sim& GIG \left( \frac{n-1}{2}+a_\eta, \phi u_1, \phi u_3 + 2 b_\eta \right),\\
\pi(\phi|\ldots) &\sim& Ga \left( \frac{n-1}{2}+a_\phi,\frac{u_1}{2\eta} - u_2 + \frac{u_3 \eta}{2} \right), \\
\pi(v_i|\ldots) &\sim & GIG \left( -1, (o_i-\mu \Delta_i)^2 + \phi \eta (\Delta_i)^2 , (\beta \Delta_i)^2 + \frac{ \phi }{\eta} \right),\\
\pi(\mu,\beta|\ldots) &\sim & N \left( \left[ \begin{array}{cc} \sum (\Delta_i)^2 v_i +\tau_\mu & \sum \Delta_i v_i^2 \\ \sum \Delta_i v_i^2 & \sum  v_i^2 + \tau_\beta \end{array} \right]^{-1} \left[ \begin{array}{c} \sum \Delta_i v_i o_i \\ \sum v_i^2 o_i \end{array} \right], \left[ \begin{array}{cc} \sum (\Delta_i)^2 v_i +\tau_\mu & \sum \Delta_i v_i^2 \\ \sum \Delta_i v_i^2 & \sum  v_i^2 + \tau_\beta \end{array} \right]^{-1} \right)
\end{eqnarray*}
Here $GIG$ is the Generalised Inverse Gaussian distribution, and $u_1 = \sum v_i$, $u_2=\sum \Delta_i - b_\phi$, $u_3= \sum (\Delta_i)^2/v_i$. Such a model is extremely quick and easy to fit and, from the resulting estimates of $v$, we can use an Inverse Gaussian bridge to obtain decadal estimates of climate volatilities. Writing $v(t_i,t_j)$ as the squared volatility associated with time increment $(t_i,t_j)$, the bridge result for time $t^*$ with $t_i \le t^* \le t_j$ is provided by \cite{Ribeiro2003} and gives:
\small
\begin{eqnarray}
\label{IGbridge}
\pi(v(t_i,t^*)|v(t_i,t^*)+v(t^*,t_j)=v) &\propto& v(t_i,t^*)(v-v(t_i,t^*)) \exp \left[ -\frac{\phi \mu}{2} \left\{ \frac{ (t^*-t_i)^2 }{v(t_i,t^*)} + \frac{ (t_j-t^*)^2 }{v-v(t_i,t^*)} \right\} \right],
\end{eqnarray}
\normalsize
where the quantity inside the exponent is a $\chi^2_1$ distribution. Exact methods for sampling from Equation \ref{IGbridge} are outlined in \citet{Ribeiro2003}. From the resulting estimates, we can look at the changing volatility in climate over time. Alternatively a return level plot could be created \citep[see e.g.][]{Coles2001}. Figure \ref{icecorepost} shows the posterior distributions of ($\mu$,$\beta$,$\eta$,$\phi$) and the 95\% credible intervals for volatilities $\sqrt{v}$ from fitting the above model to the ice core data after a run of 100,000 iterations, discarding 20,000 as burn-in, and thinning by 80, with all hyper-parameters $(a_\eta,b_\eta,a_\mu,b_\mu,\tau_\mu,\tau_\beta)$ set to 0.01, indicating only vague prior knowledge.\\ 

\section{Inference for marginal-data and joint-data posteriors}\label{MDPs}

In this section we distinguish between the marginal-data posterior $\pi(c_i|y_i)$ and the joint-data posterior $\pi(c|y)$, and subsequently introduce one key aspect of the modelling, finite Gaussian mixtures, that allows us to retain rich models and yet to achieve considerable algorithmic speed. We extend the ideas of the previous section, so that the NIG process is applied to climates $c$ which are now multivariate latent parameters, connected to the proxy data $y$ via a likelihood $\pi(y|c)= \int \pi(y|\theta,c) \pi(\theta|\mathcal{D})\; d\theta$. The key benefit of using mixtures of MDPs is that the highest dimensional parameter (climate $c$) can be simply marginalised out of the model and re-created at a second stage, focussing inference only on the climate volatility parameters.\\

It is first important to note the distinction between $\pi(c_i|y)$ and $\pi(c_i|y_i)$. The former is the marginalised \textit{joint} data posterior $\pi(c,v,\eta,\phi|y)$ (the left side of Equation \ref{eqn1}); the latter, our MDP, represents only the information in $c_i$ obtained from the proxy data $y_i$ at that same layer $i$. It is trivial to note that the MDP can be obtained directly from the application of Bayes' theorem to the conditionally independent likelihood:
\begin{eqnarray}
\label{MDPeqn}
\prod_i \pi(y_i|c_i) \propto \frac{\prod_i \pi(c_i|y_i)}{\prod_i \pi(c_i)}
\end{eqnarray}
There are two features of our model which make the use of such MDPs feasible. First, we are happy to cut feedback so that the parameters $\theta$ are integrated out of the likelihood. Second the prior on climate is intrinsic so, provided that a flat prior is applied to $\pi(c_1)$, all other $\pi(c_j)$ are flat too, thus the denominator of Equation \ref{MDPeqn} is a constant. Now the creation of $\pi(c_i|y_i)$ for every layer can be achieved in parallel at an offline early step of the process, with the lookup of $\pi(c_i|y_i)$ now being near-instantaneous. The `cutting feedback' from likelihood parameters is an unremarkable and implicit assumption in most studies involving instrumental data; e.g., experimental data on temperature are not normally used to re-calibrate the thermometer that generated them. It is interesting to note that in many branches of applied statistics what is considered `data' might actually be an MDP, though Equation \ref{MDPeqn} would never be explicitly considered. \\

More formally, we now consider replacing the likelihood in Equation \ref{eqn1} with the \textit{product MDP} $\prod_i \pi(c_i|y_i)$ which we write as $\pi_{\mathrm{MDP}}(c|y)$ to avoid confusion with the marginalised joint data posterior $\pi(c|y)$. Assuming the above requirements are met (i.e. intrinsic prior on $c$, and feedback cut between likelihood parameters $\theta$ and $y$) there are several further advantages to performing inference with the MDP: 
\begin{itemize}
\item Once $\pi(c_i | y_i)$ has been created for every $i$, we never have to make another call to a potentially expensive product likelihood
$\prod_{i=1}^{n} \pi(y_{i}|c_{i})$.
\item $c$ is of lower dimension than $y$, and thus is easier to store in large quantities.
\item Different methods for creating the marginal-data posterior can be compared without any need to re-run the full likelihood computation. Although in this paper we use MDPs constructed by our research group, the methodology we describe can in principle receive input from any Bayesian procedure.
\item We can explore marginal-data posterior distributions, and remove them at will to cross-validate.
\end{itemize}

In fact, such an approach may have advantages in other areas of Bayesian statistics. Clearly the approach does not require a flat prior on $c_1$, although that is what we use: any prior distribution in which the joint and marginal prior distributions are available will yield a suitable joint posterior. We now elaborate with a simple case in which the likelihood, and thus also the MDP, is Gaussian.

\subsection{A simple example}

For illustration we consider a simple case with fixed equal time steps and a simple Gaussian random walk prior $c_i-c_{i-1} \sim N(0,v)$. We assume a likelihood $y_i|c_i \sim N(c_i, \tau_i^{-1})$ with known $\tau_i$ so the MDP is trivially $c_i|y_i \sim N(y_i, \tau_i^{-1})$. The target of the estimation is $\pi(c,v|y)$. Some simple algebra shows that the posterior factorises: $\pi(c,v|y) \propto \pi_{\mathrm{MDP}}({c}| {y}) \pi({c}|v) = \pi({c}| {y}, v) \pi({y}|v) \pi(v)$. The distribution $\pi({c}| {y}, v)$ here is Gaussian: ${c}| {y}, v \sim N({c}; VDy, V)$. The distribution $\pi(y|v)$ is also Gaussian on first differences of $y$ (though this does not extend to the more complicated examples we present in the next section).  Here $D$ is the diagonal precision matrix with known components $\tau_i$ and $V = (D + W)^{-1}$. $V$ in turn involves the precision matrix $W$ of the random walk, being $v^{-1}B^TB$ where $B$ is an $(n-1)\times n$ differencing matrix with the first row structured as $(-1,1,0 ,\ldots, 0)
$, and subsequent rows structured similarly. Note that despite the prior on $c$ being improper all matrices involved in $\pi(c|y,v)$ and $\pi(y|v)$ are of full rank. The precision matrix $(D+W)$ is itself tri-diagonal and so allows fast inversion. Taken altogether, we can design a fast algorithm for computing the posterior by first performing inference on the parameter $v$ (by marginalising over $c$) and subsequently creating a posterior for $c$ via $\pi(c|y,v)$.\\

\subsection{Multivariate Finite Mixtures for MDPs}

The model above can be extended to the non-linear case where the MDP is $c_i|y_i \sim N(\mu_i=\mu(y_i),d_i^{-1}=d(y_i)^{-1})$. Following the same approach we obtain the joint data posterior $\pi(c,v|y) \propto \pi({c}| {y}, v) \pi({y}|v)$, where again $c|y,v \sim N(c,VD\mu,V)$. Now, however, $\pi(y|v)$ is non-Gaussian, but the same marginalisation over $c$ is possible. We can further remove the assumption that the marginal data posterior is Gaussian by extending to the case where the MDP is treated as a finite mixture. More fully, we now consider MDPs which can be written, for an $m$-dimensional $c_i$, as $\pi( \bm{c}_i | \bm{y}_i) = \sum_{g=1}^G p_{ig} N\left(\bm{c}_i; \bm{\mu}_{ig}, \bm{\tau}_{ig}^{-1} \right)$ where we now use bold type to clearly indicate the use of multivariate climate at each layer $i$. The terms $p_{ig}$ are probabilities and  $\left(  \bm{\mu_{ig}}, \bm{\tau_{ig}^{-1}} \right)$ are vector and matrix respectively; these are all functions of the count data $\bm{y}_i$ associated with 
layer $i$. In the previous section, there was but one component in each mixture and we identified the sole $\mu_{i}$ with a scalar $y_i$. \\

The formulation may alternatively be expressed in terms of an underlying indicator random variable $k_i$, which with probability $p_{ig}$ selects component $g$. Conditional on $k_i$, the MDP is simply Gaussian, and we have but a small extension of the previous section. We propose this as an approximation to the true MDPs. In the Bayesian situations we envision, the $\pi( \bm{c}_i | \bm{y}_i)$ are typically only available through (Markov Chain) Monte Carlo samples. The mixtures are then derived by fitting a finite Gaussian mixture to each sample. As detailed in Section \ref{previous}, we use the MClust algorithm of \cite{Fraley2002}. This is a once-only exercise; there is no inference in the following on any of these parameters, which are taken to be known. The implications of the approximation are that we have access to the very simple results in the previous section.\\

It follows that
\begin{eqnarray}
\pi_{\mathrm{MDP}}(\bm{c}|\bm{y}) &=& \prod_i \pi( \bm{c}_i | \bm{y}_i) = \prod_i \left\{
                \sum_g p_{ig} N\left(\bm{c}_{ig}; \bm{\mu}_{ig}, \bm{\tau}_{ig}^{-1}  \right) \right\}
            =   \sum_{\bm{K}} p_{\bm{K}} N\left(\bm{c}; \bm{\mu}_{\bm{K}}, \bm{\tau}_{\bm{K}}^{-1}  \right)
\end{eqnarray}

Here the index $\bm{K}$ runs over n-tuples $\{k_i; i=1, \ldots n \}$ and $p_{\bm{K}} = \prod_{i, g_i } p_{ig}$. If each mixture involves $G$ components, then there are $G^n$ versions of $\bm{K}$; the vectors $\bm{\mu}_{\bm{K}}$ hold the appropriate stacked  $\bm{\mu}_{ig}$ and the block diagonal matrices $\bm{\tau}_{\bm{K}}$ have blocks $\bm{\tau}_{ig}$. We abuse notation somewhat for simplicity, allowing the subscripts $(ig)$
and $\bm{K}$ to differentiate the very differently dimensioned vectors $\left( \bm{\mu}_{ig}, \bm{\mu}_{\bm{K}}\right)$ and matrices $\left( \bm{\tau}_{ig}, \bm{\tau}_{\bm{K}}\right)$. Again ${\bm{K}}$ is an index vector, with associated probability $p_{\bm{K}}$. All terms are known. It is to be recalled that the above refers simply to the product of MDPs; it is not the joint distribution $\pi( \bm{c}| \bm{y})$ that we seek here.\\

In the following sections we replace the likelihood corresponding to the observation equation with the product MDP represented as mixtures of Gaussians. By integrating out $c$ we thus bypass the evolution equation and are able to focus our inference on the volatilities at a first stage, updating them via standard Metropolis Hastings MCMC. Subsequently, we can produce climates $c$ via simple mixtures of multivariate normal distributions defined by the MDPs and the posterior distribution of the volatilities. \\

\section{Palaeoclimate modelling using MDPs}\label{model}

To complete our palaeoclimate model, we combine the MDP approach presented above, which allows us to marginalise over climate $c$, with the NIG prior used in Section \ref{motivating}. The NIG prior is applied to each climate dimension individually, so that the time series are treated as independent \textit{a priori}, though dependence is induced through the MDPs which capture relationships between the climate variables. At this stage we also include our age-depth model which allows for uncertainty in the chronology measurement. We use Bchron \citep{haslettparnell2008b} to provide very many samples of $t_1,\ldots,t_n$, each with the monotone restriction $t_i-t_{i-1}>0$. In fact, the Compound Poisson Gamma Process used in Bchron could be considered another intrinsic prior and so might be updated as part of our model. However, we do not explore that approach further here, instead preferring to again cut feedback between the age-depth model and the other parameters.\\

In summary, the following analysis stages are required to perform inference on the joint posterior distribution of climate and volatility:
\begin{enumerate}
\item Creation of a suitable chronology via, e.g. Bchron.
\item (P) Creation of MDPs for each layer in the core.
\item (P) Approximation of MDPs as Gaussian mixtures.
\item MCMC run to estimate volatilities $\bm{v}$.
\item Posterior creation of climate $\bm{c}$
\item Interpolation of volatilities $\bm{v}$ and climates $\bm{c}$ on a pre-chosen grid
\end{enumerate}
This section outlines steps 4, 5 and 6. The inclusion of (P) indicates that this stage can be performed in parallel. The full set of parametric relationships are shown in the DAG of Figure \ref{DAG}.\\

Our final model can be written as:
\begin{eqnarray}
\bm{c}(t_i)|\bm{y}_i \sim \sum_{g=1}^G p_{ig} N(\bm{\mu}_{ig},\bm{\tau}_{ig}^{-1}),\; c_j(t_i) - c_j(t_{i-1}) \sim N(0,v_{ij}),\; v_{ij} \sim IG_2(\eta_j\Delta_{i},\phi_j\Delta_{i}),
\end{eqnarray}
where all notation is as before except for the subscript $j$ denoting climate dimension from 1 to $m=3$. We require the posterior:
\begin{eqnarray}
\pi(\bm{c},\bm{v},\bm{\eta},\bm{\phi},\bm{K},\bm{t}|\bm{y},\bm{d}) &\propto&  \left[ \prod_{j=1}^m \pi(\eta_j) \pi(\phi_j) \right] \times \pi(\bm{K}) \times \pi(\bm{t}|\bm{d}) \times \left[ \prod_{j=1}^m \prod_{i=1}^{n-1} \pi(v_{ij}|\eta_j,\phi_j,\Delta_i) \right]\nonumber \\ 
& & \times \left[ \prod_{j=1}^m \prod_{i=1}^{n-1} \pi(c_j(t_{i+1})|c_j(t_{i}),v_{ij}) \right] \left[ \prod_{i=1}^n \pi_{\mathrm{MDP}}(\bm{c}(t_i)|\bm{y}_i,k_i) \right]
\end{eqnarray}

After some algebra (similar to that in Section \ref{MDPs}) we obtain:
\begin{eqnarray}
\label{fullpost}
\pi(\bm{c},\bm{v},\bm{\eta},\bm{\phi},\bm{K},\bm{t}|\bm{y},\bm{d}) &\propto& A \prod_{j=1}^m \frac{ N(\bm{c}_j;\bm{V}_{j\bm{K}} \bm{D}_{j\bm{K}} \bm{\mu}_{j\bm{K}},\bm{V}_{j\bm{K}}) N(\bm{0};\bm{\mu}_{j\bm{K}},\bm{D}_{j\bm{K}}^{-1}) }{ N(\bm{0};\bm{V}_{j\bm{K}} \bm{D}_{j\bm{K}} \bm{\mu}_{j\bm{K}},\bm{V}_{j\bm{K}}) }
\end{eqnarray}
\normalsize
with $A = \pi(\bm{\phi},\bm{\eta},\bm{K},\bm{t}|\bm{d}) \left[ \prod_{j=1}^m \prod_{i=1}^{n-1} \pi(v_{ij}|\eta_j,\phi_j,\Delta_j) \right] \left[ \prod_{j=1}^m \prod_{i=1}^{n-1} \left(v_{ij}\right)^{-1/2} \right]$, $\bm{\mu}_{j\bm{K}} = (\mu_{1jk_1},\ldots,\mu_{njk_n})  $, $\bm{D}_{j\bm{K}} = diag(\tau_{1jk_1},\ldots,\tau_{njk_n})$, $\bm{V}_{j\bm{K}} = ( \bm{D}_{j\bm{K}} +\bm{W}_j )^{-1}$ and $\bm{W}_j = \sum_i v_{ij}^{-1} \bm{B}_i \bm{B}_i^T$ where $\bm{B}_i$ is the $i$th row of difference matrix $\bm{B}$. Thus the posterior again factorises and we are left with a likelihood containing the parameters of interest $\bm{\phi,\eta,v}$. Once inference on these has been performed, we can then obtain $\bm{c}|\bm{y},\ldots$ from:
\begin{eqnarray*}
\pi(\bm{c}|\bm{y},\ldots) = \int \pi(\bm{c}|\bm{K},\bm{t},\bm{v},\bm{y},\bm{d}) \pi(\bm{K},\bm{t},\bm{v},\bm{\phi},\bm{\eta}|\bm{y},\bm{d}) \; d(\bm{K},\bm{t},\bm{v},\bm{\phi},\bm{\eta})
\end{eqnarray*}
where the first term in the integrand is a normal distribution (given by the left hand side of the numerator of Equation \ref{fullpost}) and the second is the posterior from the first inference stage.\\

We leave the computational details of fitting the model defined in Equation \ref{fullpost} to Technical Appendix \ref{MCMC}. It suffices to say that the parameters $\bm{v}$ require Metropolis-Hastings updates whilst the complete conditionals for $\bm{\eta}$ and $\bm{\phi}$ are available as Generalised Inverse Gaussian and Gamma distributions respectively. In fact we find it convenient to fix the values of $\bm{\eta}$ and $\bm{\phi}$ to some suitable prior values. The computation time is substantially reduced because $\bm{V}=(\bm{D}+\bm{W})^{-1}$ is tri-diagonal and thus requires only $\mathcal{O}(n)$ inversion steps. Similarly, for a new proposed value of $v_{ij}$, say $v_{ij}^*$, we can write $\bm{V}^* = (\bm{V}^{-1} + \left( (v_{ij}^*)^{-1}-v_{ij}^{-1} \right) \bm{B}_i \bm{B}_i^T )^{-1}$, where $\bm{B}_i$ is the $i$th row of $\bm{B}$. Inversions of $\bm{V}^*$ can be computed via the fast Woodbury formula \citep[e.g.][]{Press2002}. Once this inference stage is completed, we can perform draws of climate $c$ 
from the predictive distribution, and subsequently interpolate using the Inverse Gaussian bridge (as in Section \ref{motivating}) to estimate $\bm{v}$ at intermediary time points. Interpolations for $\bm{c}$ can be found from the resulting Brownian Bridge.\\

We incorporate time uncertainty in our algorithm by sampling a chronology from $\pi(t|d)$ at every iteration. This involves another cutting feedback assumption as the parameters of the chronology model, and hence the set of times $t$, are informed only by the radiocarbon dates in the core and not by any information on the sedimentation from the pollen. We feel that this is reasonable though it is an area for further research given that the pollen production of certain varieties of plants or tree surrounding a lake may inform the sedimentation process. The general inclusion of time uncertainty in our model greatly reduces our ability to discern precise volatilities, as can be seen in our case study below. We discuss this property further in Section \ref{discuss}.\\

In Technical Appendix \ref{validation}, we perform a number of validation steps to discern the performance of our model in ideal and non-ideal simulated situations. As the climate change prior (the evolution equation) is our main focus here and not the likelihood, we use a number of simplified versions of the full likelihood to run tests (though see Appendix \ref{validation} for the full description). We use the following steps to test the algorithm:
\begin{enumerate}
\item Fix parameters $\eta_j$ and $\phi_j$ for $j=1,2,3$.
\item Generate $v_{ij}$ from the $IG(\eta_j,\phi_j)$ for $i=1,\ldots,n-1$ and $j=1,2,3$.
\item Generate climates $c_{ij}$ from the evolution equation.
\item Generate `psuedo pollen' from $f_\theta$ with $\theta$ fixed.
\end{enumerate}
We then perform the steps outlined at the start of this section to produce a joint posterior distribution on $c,v|y$. Note that for simplicity we assume fixed, unit time and we choose $n=100$ slices which seems reasonable given the types of pollen data we analyse. We evaluate our model by looking at 90\% and 50\% coverage of the posterior compared to the true values. We perform the above steps 1,000 times in a variety of situations where various model assumptions are violated. In Appendix \ref{validation} we show that, even under quite severe violations, the coverage is still extremely adequate. We perform some further model checking steps as part of the next section.

\section{Case study: Sluggan Bog, County Antrim, Northern Ireland}\label{results}

We now apply our model to a site from Northern Ireland, previously published in \citet{smithgoddard90}. Our goal is to answer our questions of interest, namely estimating the changing climate and climate volatility over time, to which we have direct access via interpolated values of $\bm{v}$ created in stage 5 of our algorithm. We infer the three climate dimensions GDD5 (length of growing season), MTCO (harshness of winter) and AET/PET (availability of moisture), as detailed in Section \ref{previous}. 28 pollen taxa were used to create MDPs via the methods of \cite{Sweeney2012}, with Bchron \citep{haslettparnell2008b} used to create an age-depth model and thus posterior distributions of the time of each pollen layer. We fix the Inverse Gaussian parameters at the posterior means from the Ice Core data in Section \ref{icecore}, so that $\eta=2.66$ and $\phi=15.33$.\\

The creation of MDPs and their approximation as mixtures is a relatively fast step taking less than 5 minutes on a modern PC with several cores, though the former is strongly dependent on the number of layers $n$. For our core we have $n=115$ layers, though other cores may have many more. The MCMC stage was run for 100,000 iterations with a burn-in period of 20,000 and thinning by 40. The resulting 2,000 iterations were checked for convergence using the R package boa \citep{Smith2005}. Posterior creation of climates, and their subsequent interpolation, are of negligible computational impact. A full run of all stages 1-6 took around 15 minutes on an Intel Core-i7 3.4GHz processor with 8 cores and 16Gb of RAM. \\

\begin{figure}[!p]
\begin{center}
\begin{tabular}{c}
\includegraphics[width=\textwidth]{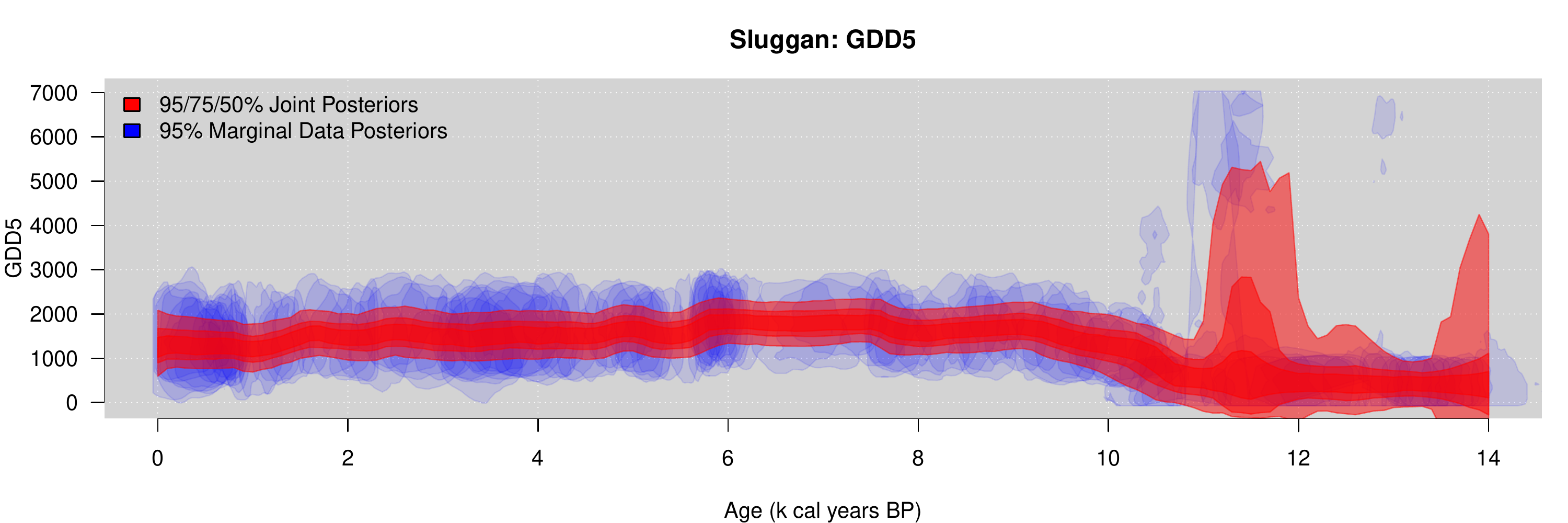}\\
\includegraphics[width=\textwidth]{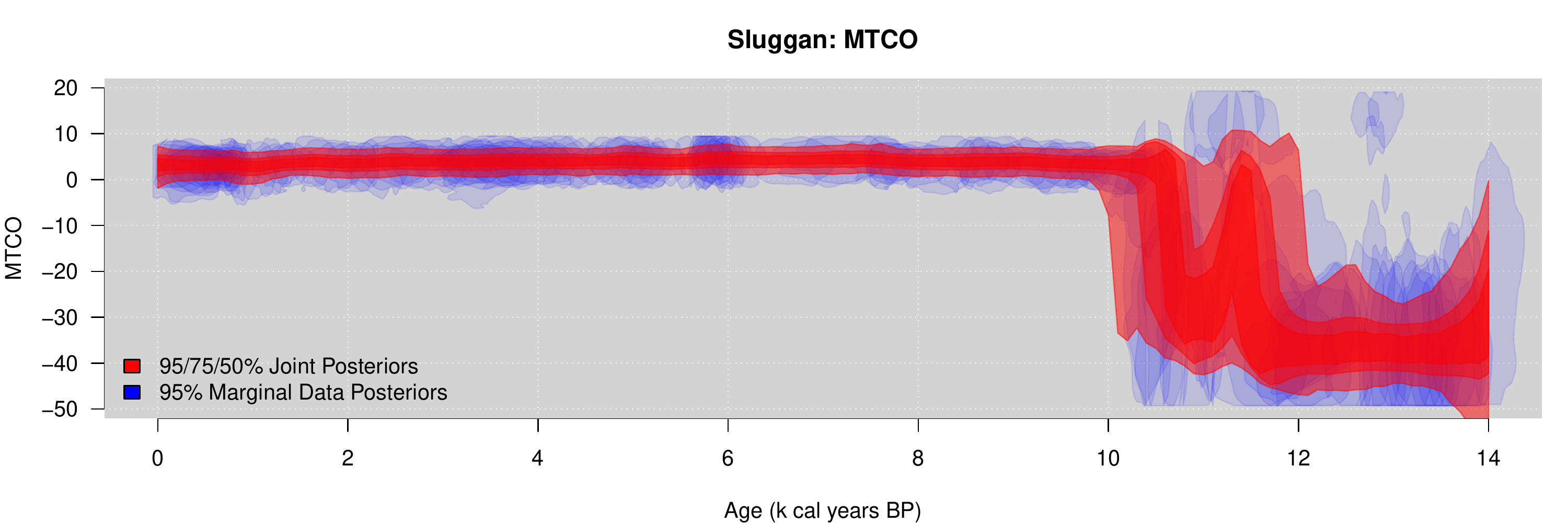}\\
\includegraphics[width=\textwidth]{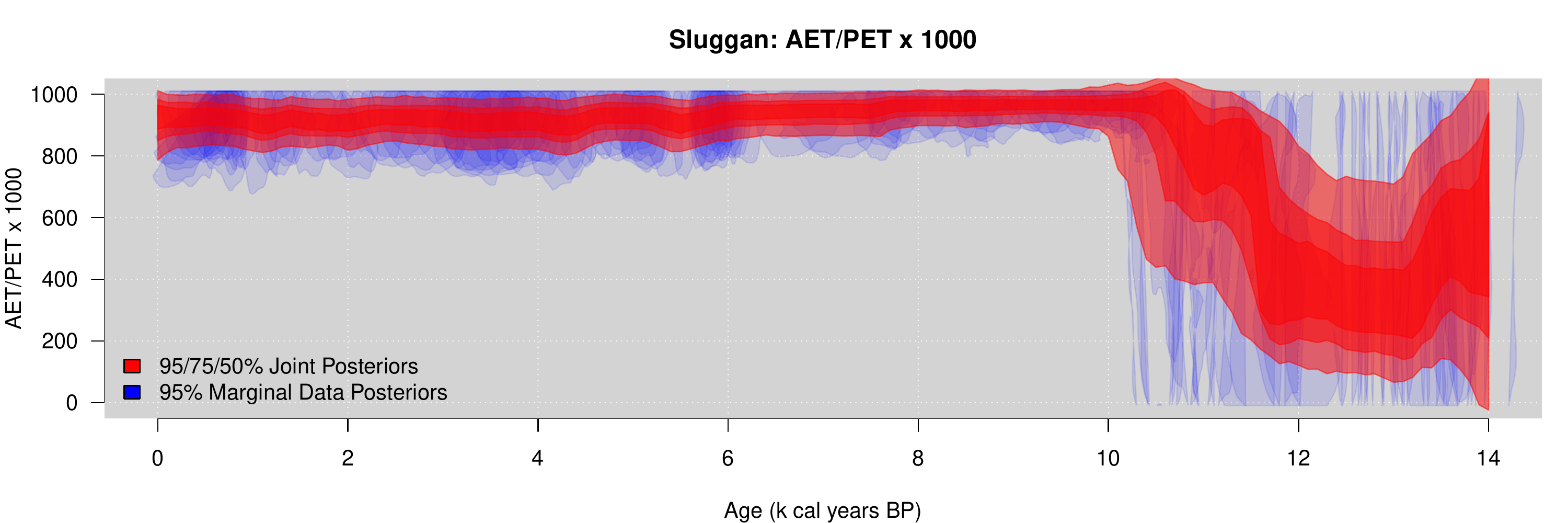}
\end{tabular}
\end{center}
\caption{
A plot of the centurial interpolated GDD5 (length of growing season), MTCO (harshness of winter) and AET/PET (proportion of available moisture; scaled up to (0,1000)) over the period 0 to 14ka BP. The blue `blobs' represent the marginal data posteriors whereas the red bands represent summarised posterior interpolations of climates $c$.}
\label{SlugganClimate}
\end{figure}

\begin{figure}[!p]
\begin{center}
\begin{tabular}{c}
\includegraphics[width=\textwidth]{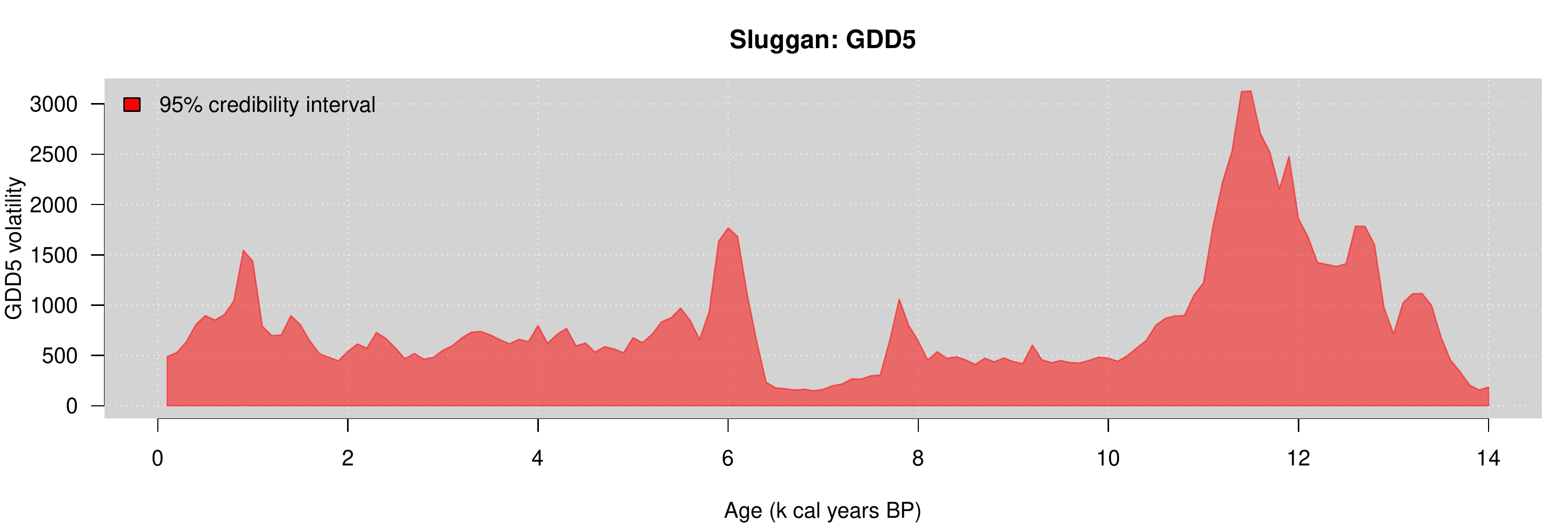}\\
\includegraphics[width=\textwidth]{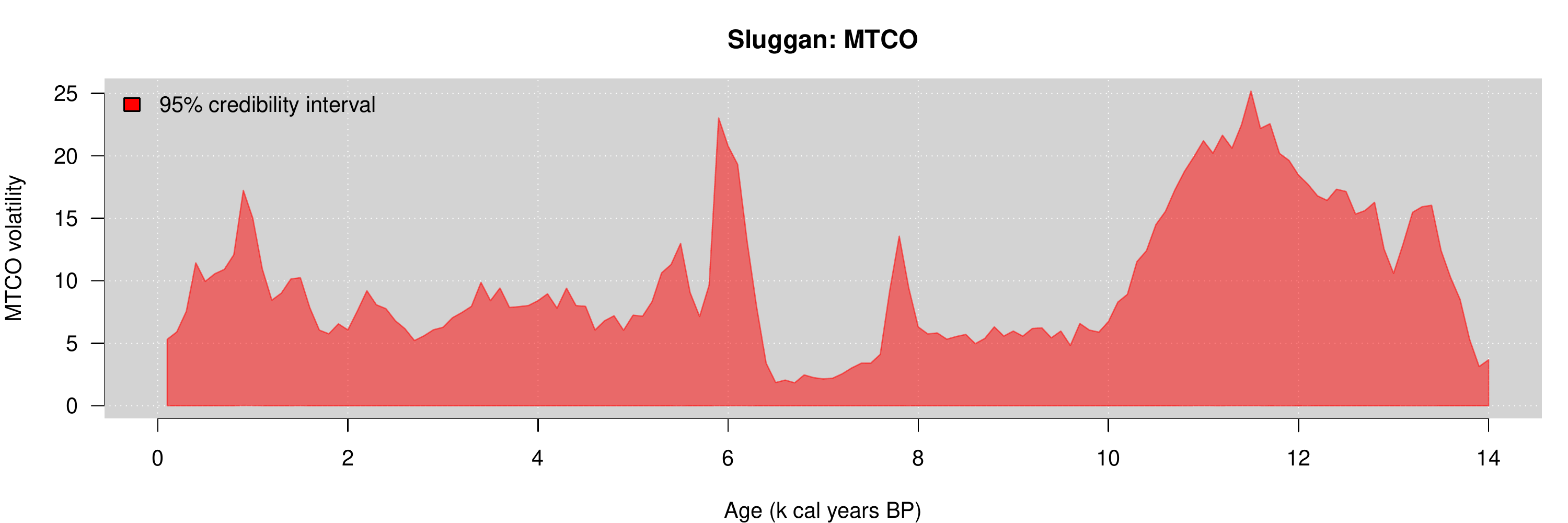}\\
\includegraphics[width=\textwidth]{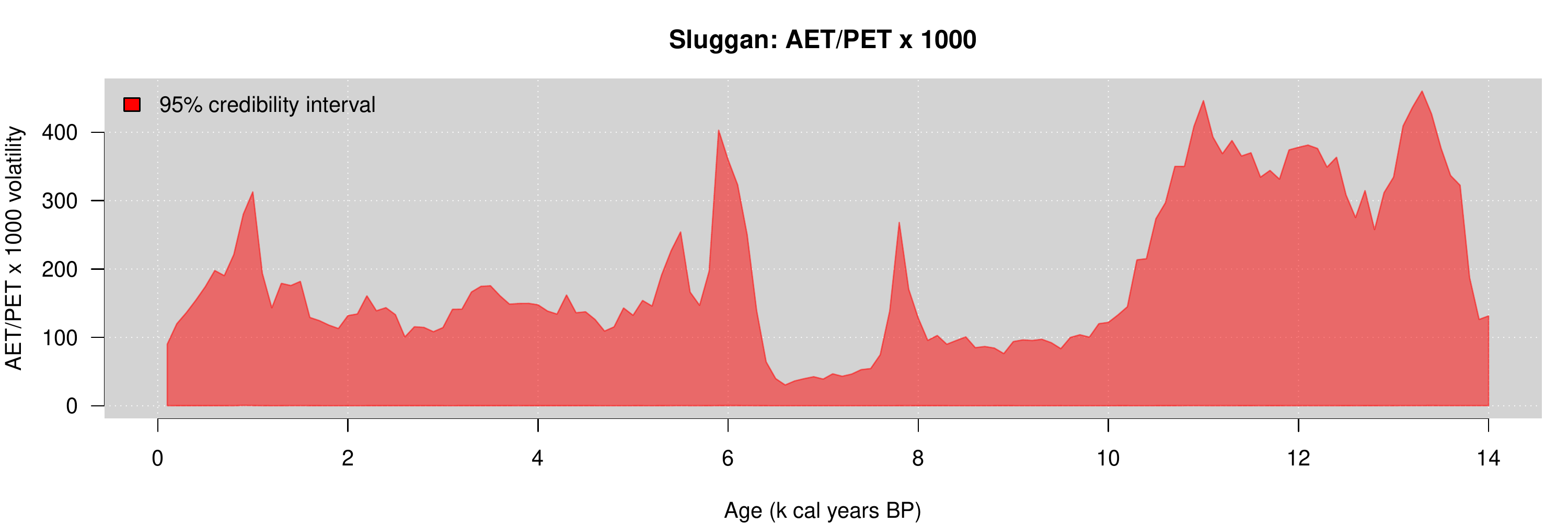}
\end{tabular}
\end{center}
\caption{
A plot of the posterior centurial interpolated volatilities for the three climate dimensions for Sluggan. The shaded areas represent 95\% credibility intervals for the centurial volatility.}
\label{SlugganVar}
\end{figure}

Figure \ref{SlugganClimate} shows GDD5, MTCO and AET/PET posterior distributions for Sluggan interpolated on to a regular centurial grid from 0 to 14 ka years BP. A Younger Dryas type event is clearly visibly in MTCO, and there appear to be similar changes in both GDD5 and AET/PET. Unsurprisingly, the last 10k years BP are reasonably constant, much like the ice core data in Figure \ref{icecore}. Figure \ref{SlugganVar} shows the posterior distributions of interpolated volatilities derived via the Inverse Gaussian bridge. The uncertainties all peak around 12ka BP, corresponding to periods of known rapid climate change. There are similar peaks at about 6ka BP, possibly corresponding to the start of agricultural activity known as the Neolithic.\\

\begin{figure}[!p]
\begin{center}
\begin{tabular}{cc}
\includegraphics[width=0.5\textwidth]{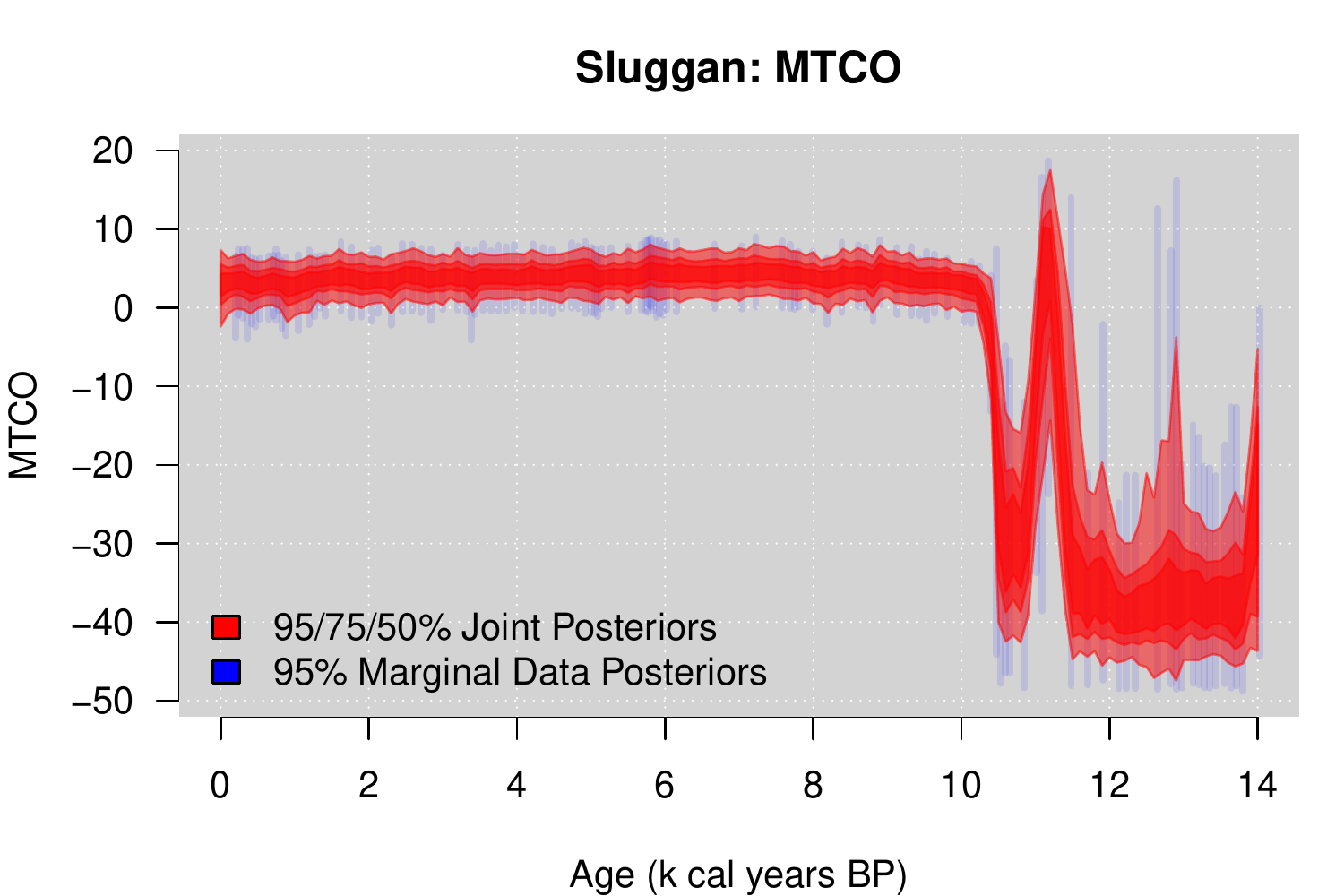} & \includegraphics[width=0.5\textwidth]{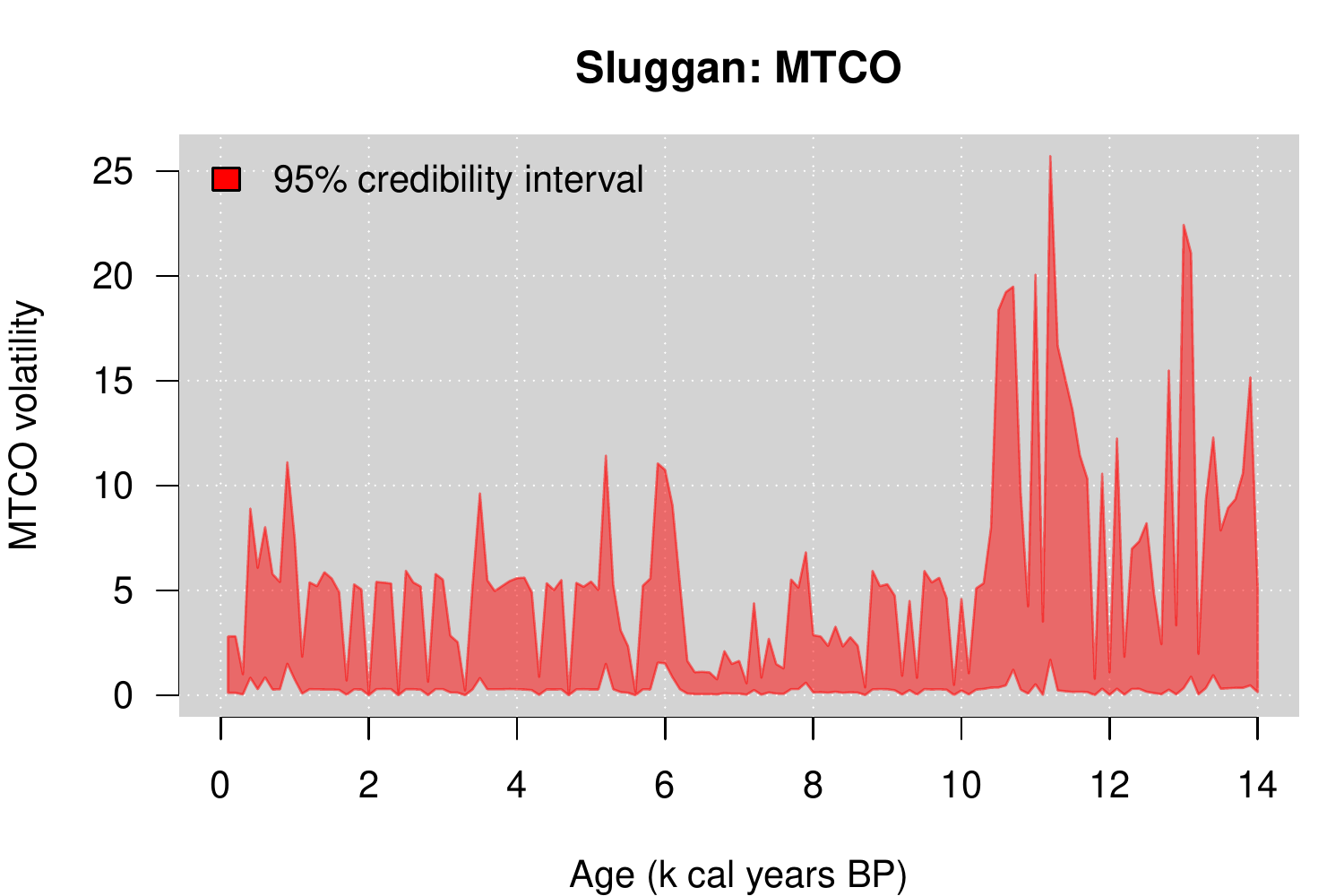}\\
\end{tabular}
\end{center}
\caption{Plots of MTCO (harshness of winter) for the Sluggan core when time is assumed fixed. The left panel shows the posterior distribution of climate $c$ whilst the right panel shows the 95\% credible intervals for the centurial interpolated volatility. The legend for the left panel is the same as that of Figure \ref{SlugganClimate}. }
\label{SlugganClimateFixedTime}
\end{figure}

\begin{figure}[!p]
\begin{center}
\begin{tabular}{cc}
\includegraphics[width=0.5\textwidth]{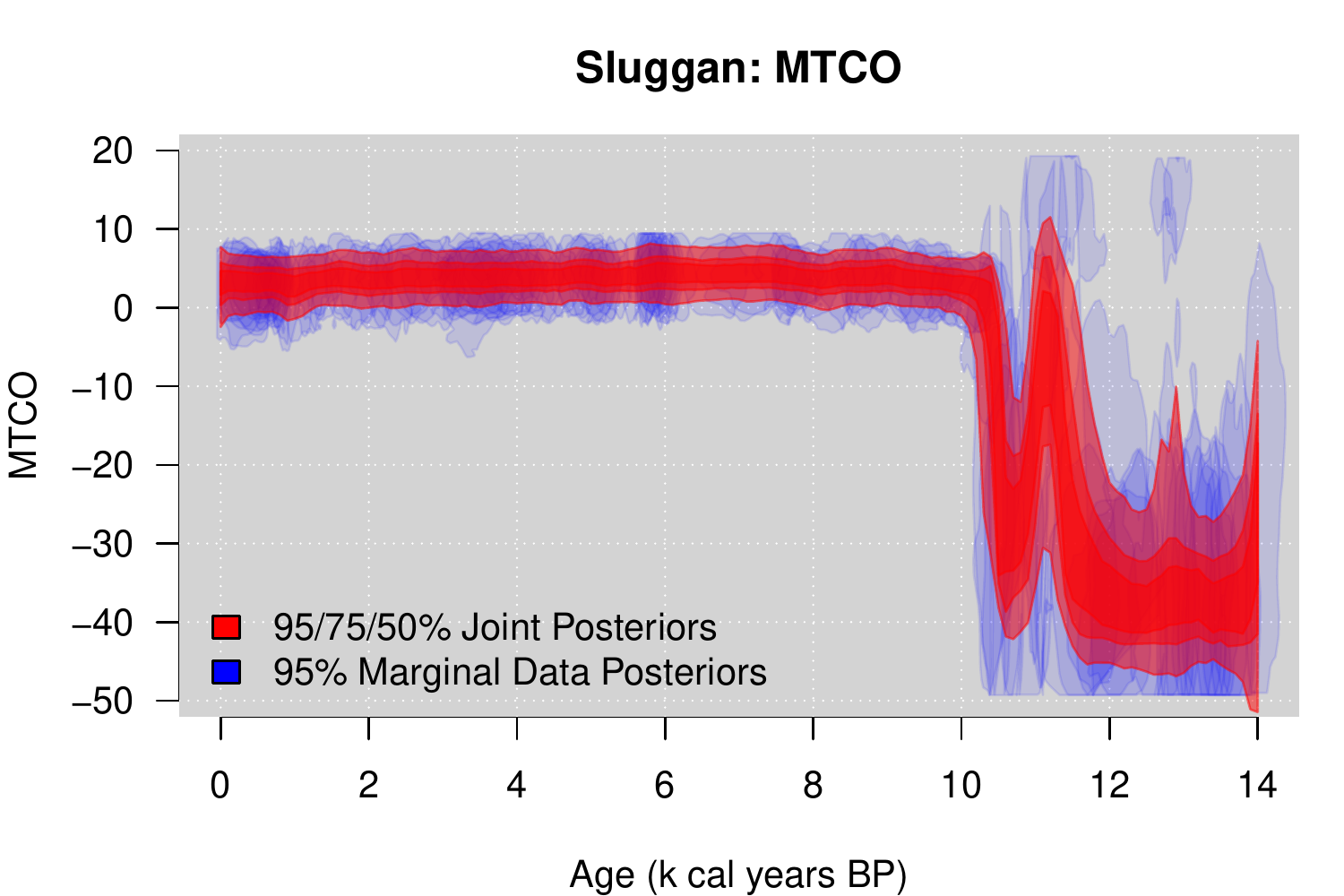} & \includegraphics[width=0.5\textwidth]{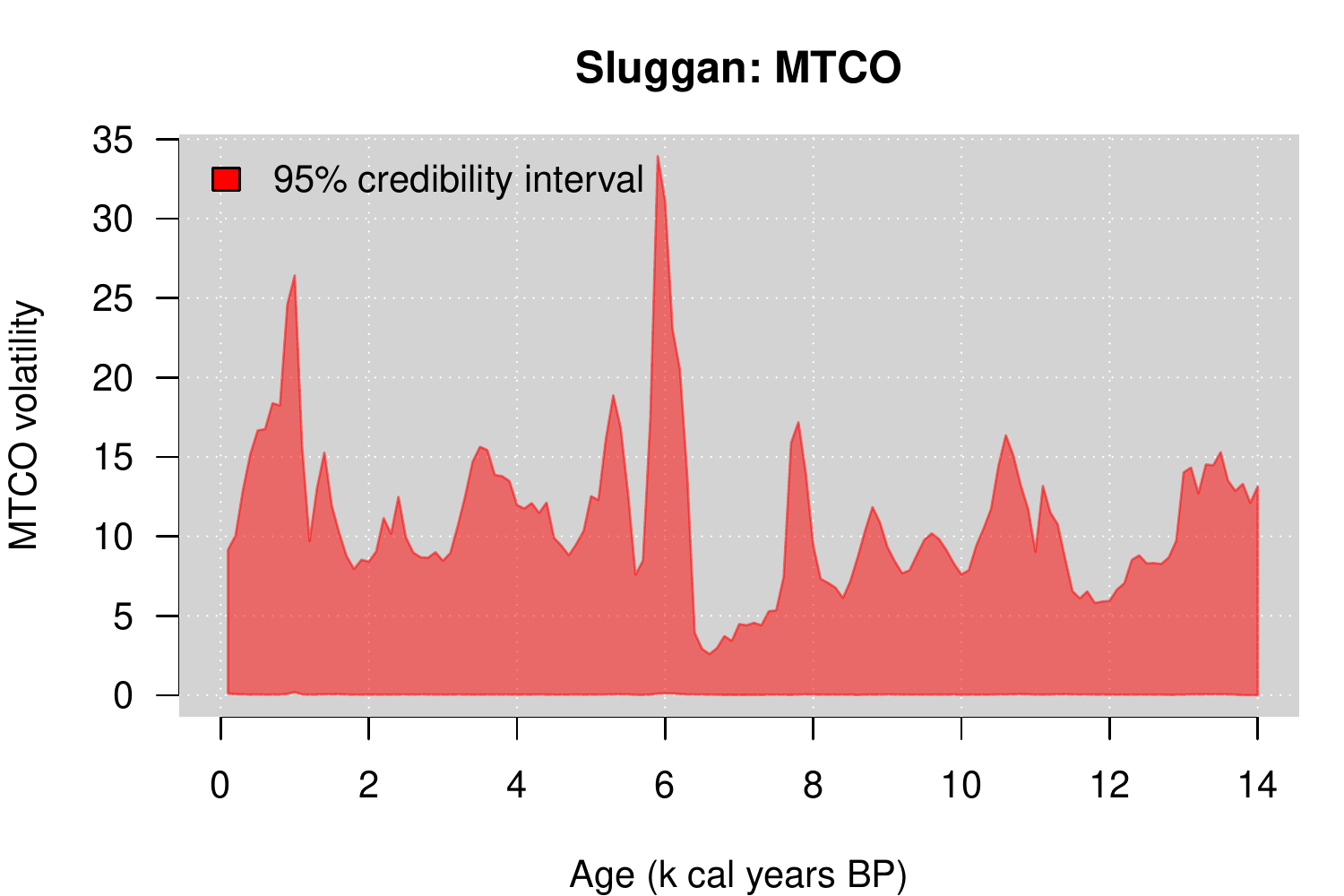}\\
\end{tabular}
\end{center}
\caption{Plots of MTCO (harshness of winter) for the Sluggan core when the evolution equation is assumed to be a Brownian motion. The left panel shows the posterior distribution of climate $c$ whilst the right panel shows the 95\% credible intervals for the centurial interpolated volatility. The legend for the left panel is the same as that of Figure \ref{SlugganClimate}. }
\label{SlugganClimateBM}
\end{figure}

It is important to note that the volatilities in Figure \ref{SlugganVar} are given as 95\% credibility intervals. These show that we have large uncertainty about the volatility which is unsurprising given that this is an estimate of a latent process with temporal uncertainty. In Figure \ref{SlugganClimateFixedTime}, we show one of the climate dimensions (MTCO) with time uncertainty ignored. The MDPs are now vertical lines as they no longer exhibit any `horizontal' uncertainty. Unsurprisingly the volatilities are somewhat more clearly defined. In Figure \ref{SlugganClimateBM}, we show MTCO with a Gaussian instead of NIG random walk (equivalent to letting $\phi \rightarrow \infty$). Note that, whilst the large change in climate around 11k years BP is still shown, it is associated with much lower volatility and thus a slower climatic change.\\

\section{Discussion}\label{discuss}

The model we have presented performs inference on palaeoclimate in a more detailed and thorough fashion than previously possible. The foundation of the model is a state space formulation which explicitly separates out the dynamical system (climate; the evolution equation) from the forward model (the link between climate and proxy pollen data; the state equation). This idea, proposed originally in \citet{haslettetal2006}, had also been suggested by \citet{Tingley2012}. We have implemented and considerably expanded this approach and developed a fast algorithm which can perform inference on both climate and climate volatility through the use of mixtures of marginal data posteriors.\\

There are several drawbacks to the model as proposed. First, it is conceivable that the mixture indicators do not cover a sufficient variety of tuples $\bm{K}$ to sufficiently learn the climate volatility parameters. Such a problem will increase with $n$ and $G$ the number of mixture components used. However, our model validation steps show that this is almost never the case and coverage properties, even when $G$ is under-estimated, still seem adequate. Another disadvantage is the cutting feedback assumptions, first between the likelihood parameters $\theta$ and the rest of the model, and second between the chronology model and the NIG process. The former seems most reasonable, as new cores are unlikely to impact much on the climate process given the strength of modern analogue data available. The latter, however, poses an interesting challenge, as if the sedimentation process is also posed as an intrinsic prior it is feasible for inclusion in our MDP-style inferential approach.\\

The modularity invoked by following our modelling assumptions is of interest in future extensions. This modularity enables various steps to be run in parallel, and also allows us to change modules as required. The steps for the different modules are given at the start of Section \ref{model}. For example, to produce the fixed time graph in Figure \ref{SlugganClimateFixedTime}, steps 2 and 3 (creation of MDPs and their approximation as mixtures) did not need to be re-run, being entirely independent of any chronological uncertainty. Thus only the other steps were required. This has larger implications for future modelling as, for example, a new forward model can be used in place of the one we use with no other changes required to any of the other steps. The same applies for the chronology model, the Mclust mixture algorithm, and the evolution equation itself (though it would still need to be intrinsic).\\ 

There are several enhancements which follow immediately and to which our algorithm may well still apply. These include: multi-proxy inference, development of new forward models, and spatio-temporal approaches:
\begin{itemize}
\item A multi-proxy analysis of palaeoclimate would require multiple forward models describing the relationship between climate and the various proxies. Our state-space MDP approach would not be hindered by such an extension, as we could simply create MDPs for the different proxies and include them as standard, so that the product MDP now becomes, e.g. $\pi_{\mathrm{MDP}}^{\mathrm{proxy1}}(c|y_1) \times \pi_{\mathrm{MDP}}^{\mathrm{proxy2}}(c|y_2)$. However, care needs to be taken in selecting the aspects of climate to which the different proxies are responsive. If these are substantially different, it may be that an extra layer of the state space model is required to match the different climate variables appropriately.
\item A key development in palaeoclimate will be the advancement of probabilistic forward models describing the causal chain from climate to proxy data. The forward model we use is relatively simplistic in its description of the mechanics of climate/pollen interaction, though it is far more sophisticated in its description of the uncertainty relating to the counting of pollen data and the relationships between pollen varieties. We encourage the development of physical forward models provided they retain suitable stochastic elements. A recent example of such thinking is \cite{Tolwinsk2011}.
\item Finally, we might extend our state space approach into the spatial domain and give a richer evolution equation:
\begin{eqnarray*}
\label{statespace3}
y(s_i,t_i)|c(s_i,t_i) \sim f_\theta(c(s_i,t_i)),\; i=1\ldots,n\\ 
\label{statespace4}
c(s_i,t_i)|\left\{ c(s_1,t_1),\ldots,c(s_{i-1},t_{i-1}) \right\} =c \sim \zeta_\kappa(c), \; i=2,\ldots,n
\end{eqnarray*}
where now both pollen and climate are indexed by space $s$ and time $t$ and the prior distribution $\zeta$ is applied to climate change, parameterised by $\kappa$. We might assume that this would use all observations from previous time points $t_1,\ldots,t_{i-1}$ so that a particle algorithm might now become more appropriate. The prior $\zeta$ might be a simple stochastic climate model, or a richer version of our random walk including covariates and a spatial process. It is immediately obvious that $c$ will no longer factorise out of the posterior, yet if $\zeta$ remains intrinsic a Laplace approximation might still allow our algorithm to proceed, though with caveats as to the size of the approximation error. Finally, even in situations where the prior is not intrinsic, it may be that other non-Gaussian mixture arrangements will yield simple tractable forms. \\
\end{itemize}

Performing inference on palaeoclimate over multiple sites may be possible by following the recently proposed methodology of \cite{Lindgren2011}. In fact, the borrowing of strength from nearby cores may overcome one of our main issues: that of temporal uncertainty. It is certainly feasible that the constrained correlation of neighbouring sites will reduce temporal variability and thus provide more precise estimates of climate and possibly its associated volatility. Following this approach seems most promising in producing a pan-European map of palaeoclimate and its uncertainty.\\

\section{Acknowledgements}

We would like to thank the following for fruitful discussions as part of the SUPRAnet network: Jonty Rougier, Caitlin Buck, Tamsin Edwards, and Michel Crucifix. We would also like to thank Brendan Murphy for his assistance with the mclust algorithm.

\bibliography{/Users/andrewparnell/Dropbox/bibtex/library}
\bibliographystyle{chicago}

\appendix

\newpage
\section{Notation}\label{notation}

Below is a table outlining the notation we use throughout the paper and especially in the mathematical derivations contained in Appendix \ref{MCMC}.

\begin{tabular}{lp{14cm}}
$c_{ij}$ & $=c_j(t_i)$ latent palaeoclimate at time $t_i$ for climate dimension $j$\\
$\bm{c}$ & vector (length $m \times n$) of latent climate $c_{ij}$ at all times and in all dimensions. Sorted first by time and then by dimension\\
$\bm{c}_i$ & $=\bm{c}(t_i)$ vector (length $m$) of latent palaeoclimate where each element represents each climate dimension at time $t_i$\\
$\bm{c}_j$ & vector (length $n$) of latent palaeoclimate where each element represents climate over all time for climate dimension $j$\\
$d$ & Depths\\
$f$ & Forward model \\
$g$ & group number/mixture component\\
$i$ & layer number\\
$j$ & climate dimension number \\
$k$ & index number\\
$k_i$ & index number for layer $i$ taking values 1 to $G$\\
$m$ & Total number of climate dimensions \\
$n$ & Total number of layers \\
$o_i$ & $\delta^{18}$O measurement at time $t_i$ \\
$p_{ig}$ & probability of mixture component $g$ for layer $i$\\
$t_i$ & time of layer $i$ in thousands of years before present (ka BP)\\
$v_{ij}$ & variance of random walk for layer $i$, climate dimension $j$\\
$\bm{y}$ & pollen data\\
$\bm{y}_i$ & vector of pollen data for layer $i$\\
$z_i$ & $n$-vector of zeros with 1 in the $i$th and -1 in the $(i+1)$th position\\ 
\end{tabular}

\begin{tabular}{lp{14cm}}
$\bm{D}_{j\bm{K}}$ & Diagonal matrix of precisions for climate dimension $j$ under mixture components $\bm{K}$\\
$G$ & Total number of mixture components\\
$\bm{I}_n$ & Identity matrix of size $n$\\
$\bm{K}$ & Index set of $k_1,k_2,\ldots,k_n$\\
$\bm{V}_{j\bm{K}}$ & Variance matrix for climate dimension $j$ under mixture components $\bm{K}$. $\bm{V}_{j\bm{K}} = (\bm{D}_{j\bm{K}}+\bm{W}_j)^{-1}$ \\
$\bm{W}_j$ & Random walk precision matrix for climate dimension $j$\\
\end{tabular}

\begin{tabular}{lp{14cm}}
$\beta$ & Skew parameter for Inverse Gaussian distribution \\
$\Delta_i$ & Time differences $t_i - t_{i-1}$\\
$\eta$ & Parameter of Inverse Gaussian Distribution \\
$\theta$ & Set of parameters governing forward model\\
$\mu$ & Mixture component means, also drift parameter for ice core model\\
$\mu_{ijg}$ & Mixture mean for layer $i$, climate dimension $j$, mixture component $g$\\
$\bm{\mu}_{ig}$ & $m$-vector of mixture means for layer $i$ when mixture component $g$ is chosen \\
$\bm{\mu}_{j\bm{K}}$ & $n$-vector of mixture means for all layers on climate dimension $j$ under mixture components $\bm{K}$\\
$\pi$ & probability distribution\\
$\tau$ & Mixture component precisions \\
$\tau_{ijg}$ & Mixture precision for layer $i$, climate dimension $j$, mixture component $g$\\
$\bm{\tau}_{ig}$ & $m\times m$-matrix of mixture precisions for layer $i$ when mixture component $g$ is chosen - often diagonal \\
\end{tabular}

\newpage
\section{MCMC details}\label{MCMC}

We require posterior simulations of $\bm{v}$ via the Metropolis-Hastings algorithm; $\bm{c}$ now having been integrated out and hyper-parameters $\bm{\phi},\bm{\eta}$ are treated as fixed. The complete conditional for $v_{ij}$ is independent of the other $v$'s so below we drop the subscripts $j$ for clarity. We update $v$ by proposing new value $v^{*}$, and hence create new matrix $\bm{Q_K}^{*} = \bm{Q_K}+ ((v^{*})^{-1}-v^{-1}) \bm{z}_{i} \bm{z}_{i}^{T}$ where $\bm{Q_K}=\bm{V_K}^{-1}$. Calculate MH ratio as:
\small
\begin{eqnarray*}
R_{v_{i}} &=& \frac{N(\bm{0};\bm{V_K D_K \mu_K},\bm{V_K})}{N(\bm{0};\bm{V_K}^* \bm{ D_K \mu_K},\bm{V_K}^*)} \times \frac{q(v_{i}|v_{i}^{*})}{q(v_{i}^{*}|v_{i})} \times  \frac{IG(v_{i}^{*}|\phi,\eta)}{IG(v_{i}|\phi,\eta)} \times \frac{ (v_i^*)^{-\frac{1}{2}} }{ v_i^{-\frac{1}{2}} }\\
&=& \frac{ | \bm{Q_K} |^{\frac{1}{2}} \exp \left\{ -\frac{1}{2} \left[ (\bm{V_K D_K \mu_K})^{T} \bm{Q_K} \bm{V_K D_K \mu_K} \right] \right\} }{ | \bm{Q_K}^{*} |^{\frac{1}{2}} \exp \left\{ -\frac{1}{2} \left[  (\bm{V_K}^* \bm{D_K \mu_K})^{T} \bm{Q_K} \bm{V_K}^* \bm{D_K \mu_K} \right] \right\} } \times \frac{q(v_{i}|v_{i}^{*})}{q(v_{i}^{*}|v_{i})} \times  \frac{IG(v_{i}^{*}|\phi,\eta)}{IG(v_{i}|\phi,\eta)} \times \frac{ (v_i^*)^{-\frac{1}{2}} }{ v_i^{-\frac{1}{2}} }\\
&=& \frac{ | \bm{Q_K} |^{\frac{1}{2}}}{ | \bm{Q_K}^{*} |^{\frac{1}{2}} } \exp \left\{ -\frac{1}{2} \left[ (\bm{V_K D_K \mu_K})^{T} \bm{Q_K} \bm{V_K D_K \mu_K} - (\bm{V_K}^* \bm{D_K \mu_K})^{T} \bm{Q_K}^* \bm{V_K}^* \bm{D_K \mu_K} \right] \right\} \times \frac{q(v_{i}|v_{i}^{*})}{q(v_{i}^{*}|v_{i})} \times  \frac{IG(v_{i}^{*}|\phi,\eta)}{IG(v_{i}|\phi,\eta)}\times \frac{ (v_i^*)^{-\frac{1}{2}} }{ v_i^{-\frac{1}{2}} } \times \frac{ (v_i^*)^{-\frac{1}{2}} }{ v_i^{-\frac{1}{2}} }\\
&=& \left[ \frac{ | \bm{Q_K} | }{ | \bm{Q_K}^* |} \right]^{\frac{1}{2}} \exp \left\{ -\frac{1}{2} \left[(\bm{D_K \mu_K})^T (\bm{V_K}-\bm{V_K}^*) \bm{D_K} \bm{\mu_K}) \right] \right\} \times  \frac{q(v_{i}|v_{i}^{*})}{q(v_{i}^{*}|v_{i})} \times  \frac{IG(v_{i}^{*}|\phi,\eta)}{IG(v_{i}|\phi,\eta)} \times \frac{ (v_i^*)^{-\frac{1}{2}} }{ v_i^{-\frac{1}{2}} }
\end{eqnarray*}
\normalsize
Importantly (dropping the $\bm{K}$ subscripts for the time being) $\bm{V^*}^{-1} = \bm{Q}^{*} = \bm{Q}+  \bm{z}_{i} ((v_{i}^{*})^{-1}-v_{i}^{-1}) \bm{z}_{i}^{T}$ can be manipulated via the Woodbury formulae:
\begin{eqnarray*}
(\bm{A}+\bm{UCV})^{-1} &=& \bm{A}^{-1} - \bm{A}^{-1} \bm{U} (\bm{C}^{-1}+\bm{VA}^{-1}\bm{U})^{-1}\bm{VA}^{-1}\\
|\bm{A}+\bm{UCV}| &=& |\bm{C}^{-1}+\bm{VA}^{-1}\bm{U}| |\bm{C}||\bm{A}|
\end{eqnarray*}
Thus we have:
\begin{eqnarray*}
\frac{ | \bm{Q} | }{ | \bm{Q}^{*} | } &=& \frac{  | \bm{Q} | }{ | \bm{Q}+  \Delta_i^{-1} \bm{z}_{i} ((v_{i}^{*})^{-1}-v_{i}^{-1}) \bm{z}_{i}^{T} |} \\
&=& \frac{ | \bm{Q} | }{ | \bm{Q} |  | 1 + ((v_{i}^{*})^{-1}-v_{i}^{-1}) \bm{z}_{i}^{T} \bm{V} \bm{z}_{i}|} \\
&=&  \left(1 + \left[ (v_{i}^{*})^{-1}-v_{i}^{-1} \right] \bm{z}_{i}^{T} \bm{V} \bm{z}_{i} \right)^{-1}
\end{eqnarray*}
Similarly:
\begin{eqnarray*}
\bm{V} - \bm{V}^* = \bm{Q}^{-1} -  (\bm{Q}^{*})^{-1} &=& \bm{Q}^{-1} - \bm{Q}^{-1} + \bm{Q}^{-1} \bm{z}_{i} \left( \left[ (v_{i}^{*})^{-1}-v_{i}^{-1} \right]^{-1} + \bm{z}_{i}^{T} \bm{Q}^{-1} \bm{z}_{i} \right)^{-1} \bm{z}_{i}^{T} \bm{Q}^{-1} \\
&=& \bm{Q}^{-1} \bm{z}_{i} \left( \left[ (v_{i}^{*})^{-1}-v_{i}^{-1} \right]^{-1} + \bm{z}_{i}^{T} \bm{Q}^{-1} \bm{z}_{i} \right)^{-1} \bm{z}_{i}^{T} \bm{Q}^{-1} \\
&=& \bm{V} \bm{z}_{i} \left( \left[ (v_{i}^{*})^{-1}-v_{i}^{-1} \right]^{-1} + \bm{z}_{i}^{T} \bm{V} \bm{z}_{i} \right)^{-1} \bm{z}_{i}^{T} \bm{V} \\
&=& \left( \left[ (v_{i}^{*})^{-1}-v_{i}^{-1} \right]^{-1} + \bm{z}_{i}^{T} \bm{V} \bm{z}_{i} \right)^{-1} \bm{V} \bm{z}_{i} \bm{z}_{i}^{T} \bm{V}
\end{eqnarray*}
Thus $R_{v_i}$ can be computed from $\bm{V}$ without the need for $\bm{V}^{*}$:
\small
\begin{eqnarray*}
R_{v_{i}} =  \left(1 + \left[ (v_{i}^{*})^{-1}-v_{i}^{-1} \right] \bm{z}_{i}^{T} \bm{V_K} \bm{z}_{i} \right)^{-\frac{1}{2}} \exp \left\{ -\frac{(\bm{D_K \mu_K})^T \bm{V_K} \bm{z}_{i} \bm{z}_{i}^{T} \bm{V_K} \bm{D_K \mu_K} }{2\left( \left[ (v_{i}^{*})^{-1}-v_{i}^{-1} \right]^{-1} + \bm{z}_{i}^{T} \bm{V_K} \bm{z}_{i} \right)} \right\} \times  \frac{q(v_{i}|v_{i}^{*})}{q(v_{i}^{*}|v_{i})} \times  \frac{IG(v_{i}^{*}|\phi,\eta)}{IG(v_{i}|\phi,\eta)} \times \frac{ (v_i^*)^{-\frac{1}{2}} }{ v_i^{-\frac{1}{2}} }
\end{eqnarray*}
\normalsize
\normalsize

Note that this involves many repeated calculations of $\bm{z}_{i}^{T} \bm{V_K} \bm{z}_{i}$ and $$(\bm{D_K \mu_K})^T \bm{V_K} \bm{z}_{i} \bm{z}_{i}^{T} \bm{V_K} \bm{D_K \mu_K} = (\bm{z}_{i}^{T} \bm{V_K} \bm{D_K \mu_K})^T \bm{z}_{i}^{T} \bm{V_K} \bm{D_K \mu_K}. $$ In \texttt{R} it is possible to use the \texttt{limSolve} package and the function \texttt{Solve.tridiag} to create $\bm{V}$ as it is the inverse of a tri-diagonal matrix. By solving the linear systems we can thus directly obtain, for example $\bm{V_K} \bm{z}_{i}$ or $\bm{V_K} \bm{D_K \mu_K}$, rather than computing $\bm{V_K}$ separately.\\

\newpage
\section{Model validation}\label{validation}

In this section we determine the properties of our model fitting algorithm using simulated data under some idealised and non-idealised circumstances. To improve the speed of our tests we simplify the likelihood somewhat, though since this is not part of our inference stage in this paper we feel this is reasonable. Similarly we simulate data only observed on fixed, unit time. We consider 5 different scenarios:
\begin{enumerate}
\item A simple Gaussian test that the parameters are identifiable when simulated from the model. We set $n=100$ and $m=3$. For $j=1,\ldots,m$ we first simulate $\eta_j \sim U(0.1,10)$ and $\phi_j \sim U(0.1,10)$. For $i=1,\ldots,n-1$ we then create $v_{ij} \sim IG_2(\mu_j,\phi_j)$ and, for $i=1,\ldots,n$, we create $c_{ij}-c_{i-1,j} \sim N(0,v_{ij})$. Finally we create $\delta_i \sim U(0.02,2)$ and simulate pseudo pollen $y_{ij} \sim N(c_{ij},\sqrt{\delta_i^{-1}})$. From the pseudo pollen data and the Gaussian likelihood we obtain Gaussian MDPs (with no simulation or mixture approximation required) which are passed, with the values of $\phi_j$ and $\eta_j$, to our MCMC functions to provide posterior distributions of 3-dimensional climate and volatility.
\item A zero-inflated Poisson likelihood with 3 pollen taxa. The IG parameters, volatilities and climates are simulated as above, but we create pseudo-pollen via $y_{i1} \sim ZIP(p_1,\sqrt{a_1c_1^2+a_2c_2^2})$ , $y_{i2} \sim ZIP(p_2,\sqrt{a_1c_1^2+a_3c_3^2})$ and $y_{i3} \sim ZIP(p_3,\sqrt{a_1c_1^2+a_2c_2^2+a_3c_3^2})$. Here $ZIP(p,r)$ is a zero-inflated Poisson distribution with zero inflation parameter $p$ and rate $r$. We set $p_1,p_2,p_3$ respectively as $p_j \sim U(0,0.2)$ and $a_1,a_2,a_3$ as Poisson rate parameters simulated as the modulus of a normal distribution: $a_j \sim |N(0,1)|$. The pseudo-pollen data are turned into MDPs via importance sampling. The ZIP model gives MDPs that are quite often multi-modal. The MDPs are then approximated as mixtures using $G=5$ mixture components. These mixture components are then passed to our MCMC algorithm to estimate climates and volatilities.
\item Exactly as (2) but using only 2 mixture components. In many situations this will be a poor representation of the MDP and thus may bias estimates of climate or volatility.
\item Split into two parts:
\begin{enumerate}
\item Exactly as (2) but with the Inverse Gaussian parameters $\eta_j$ and $\phi_j$ given an underestimating multiplicative bias value simulated from $U(0.5,1)$.
\item Exactly as (2) but with the Inverse Gaussian parameters $\eta_j$ and $\phi_j$ given an overestimating multiplicative bias value simulated from $U(1,5)$.
\end{enumerate}
\end{enumerate}
We run each of the above 1000 times, and check the coverage properties of the climate posterior to see whether they lie within the 90\% and 50\% credibility intervals. Table \ref{validationtable} shows the results. Under each scenario the model seems to perform extremely well.

\begin{center}
\begin{table}
\begin{tabular}{llll}
\hline
\hline
Scenario & Detail & Proportion& Proportion  \\
 & &  inside 90\% CI  & inside 50\% CI\\
\hline
1 & Gaussian likelihood & 90.7\% & 50.8\% \\
2 & ZIP likelihood & 90.8\% & 47.7\% \\
3 & ZIP likelihood (too few mixture components)& 90.1\% & 44.5\% \\
4a & ZIP likelihood (under-estimated IG parameters) & 91.6\% & 46.5\% \\
4b & ZIP likelihood (over-estimated IG parameters) & 94.0\% & 51.1\% \\
\hline
\hline
\end{tabular}
\caption{Performance of the different model validation scenarios}
\label{validationtable}
\end{table}
\end{center}

\newpage
\section{R Package Bclim}\label{Bclim}

Bclim is available as part of the open source, free, statistical software R \citep{RFoundationForStatisticalComputingAustria2011}. R is available to download from \url{www.r-project.org}. To install the package Bclim simply type \texttt{install.packages("Bclim")} at the R prompt. The Bclim package is made up of four main functions (covering steps 2 to 6 as given in Section \ref{model}), 2 plotting functions for (climate and climate volatility), and a function which runs all necessary steps in sequence.\\

Example data to run the function can be downloaded from \url{http://mathsci.ucd.ie/~parnell_a/Bclim.html}. To run the Sluggan example shown in Section \ref{results}, the files should be downloaded via the commands:
\begin{verbatim}
# Download and load in the response surfaces:
url1 <- 'http://mathsci.ucd.ie/~parnell_a/required.data3D.RData'
download.file(url1,'required_data3D.RData')

# and now the pollen
url2 <- 'http://mathsci.ucd.ie/~parnell_a/SlugganPollen.txt'
download.file(url2,'SlugganPollen.txt')

# and finally the chronologies
url3 <- 'http://mathsci.ucd.ie/~parnell_a/Sluggan_2chrons.txt'
download.file(url3,'Slugganchrons.txt')
\end{verbatim}
The response surfaces in the first command above are the pre-calibrated forward model parameters $\theta$. The subsequent functions use the locations of the pollen and chronology file rather than loading them into RAM:
\begin{verbatim}
# Create variables which state the locations of the pollen and chronologies
pollen.loc <- paste(getwd(),'/SlugganPollen.txt',sep='')
chron.loc <- paste(getwd(),'/Slugganchrons.txt',sep='')

# Load in the response surfaces
load('required.data3D.RData')
\end{verbatim}
The functions now proceed as \texttt{BclimLayer} which produces the marginal data posteriors, \texttt{BclimMixPar} or \texttt{BclimMixSer} which approximate the MDPs as mixtures (either in parallel or serial respectively), \texttt{BclimMCMC} which produces posterior chains of volatilities and climates, and \texttt{BclimInterp} which uses the Inverse Gaussian and Brownian bridges to interpolate climate. Finally \texttt{BclimCompile} produces a  list object which can be passed to \texttt{plotBclim} or \texttt{plotBclimVol} for plotting:
\begin{verbatim}
step1 <- BclimLayer(pollen.loc,required.data3D=required.data3D)
step2 <- BclimMixPar(step1)
step3 <- BclimMCMC(step2,chron.loc)
step4 <- BclimInterp(step2,step3) 
results <- BclimCompile(step1,step2,step3,step4,core.name="Sluggan")

# Create a plot of MTCO (dim=2)
plotBclim(results,dim=2)

# Create a volatility plot
plotBclimVol(results,dim=2)
\end{verbatim}
Each of the above functions has an associated help file which provides further information and options.

\end{document}